\documentclass[sigconf]{acmart}

\usepackage[utf8]{inputenc} % allow utf-8 input
\usepackage[T1]{fontenc}    % use 8-bit T1 fonts
\usepackage{hyperref}       % hyperlinks
\usepackage{url}            % simple URL typesetting
\usepackage{booktabs}       % professional-quality tables
\usepackage{amsfonts}       % blackboard math symbols
\usepackage{nicefrac}       % compact symbols for 1/2, etc.
\usepackage{microtype}      % microtypography
\usepackage{xcolor}         % colors
\usepackage{graphicx}
\usepackage{anyfontsize}
\usepackage{siunitx}
\usepackage{graphicx,subfigure}
\usepackage{dblfloatfix}

\usepackage{enumitem}
\usepackage{multirow}
\usepackage{balance}

\newcommand{\xref}[1]{\S\ref{#1}}

\newcommand{\squishlist}{\begin{itemize}[itemsep=1pt,parsep=2pt,topsep=3pt,partopsep=0pt,leftmargin=0em, itemindent=1em,labelwidth=1em,labelsep=0.5em]}
\newcommand{\squishend}{\end{itemize}}

\newenvironment{todo-env}{\par\color{red}}{\par}
\newenvironment{help-env}{\par\color{blue}}{\par}
\newenvironment{ready-for-review}{\par\color{violet}}{\par}

\newif\ifcomments
\commentstrue % COMMENT THIS LINE TO DISABLE COMMENTS
\ifcomments
    \newcommand{\steve}[1]{\textcolor{cyan}{[SS: #1]}}
    \newcommand{\shyam}[1]{\textcolor{magenta}{[SG: #1]}}
    \newcommand{\ira}[1]{\textcolor{red}{[I: #1]}}
\else
    \providecommand{\steve}[1]{}
    \providecommand{\shyam}[1]{}
    \providecommand{\ira}[1]{}
\fi

\title{{ClearBuds:  Wireless  Binaural Earbuds for  \\Learning-Based Speech Enhancement}}

\author{Ishan Chatterjee*, Maruchi Kim*, Vivek Jayaram*}
\author{Shyamnath Gollakota, Ira Kemelmacher, Shwetak Patel, Steven M. Seitz}
\affiliation{%
  \institution{$^*$Co-primary student authors}
    \institution{Paul G. Allen School of Computer Science \& Engineering}
  \institution{University of Washington, Seattle, WA}
  \country{USA}
}
\email{{ichat, mkimhj, vjayaram, gshyam, kemelmi, shwetak, seitz}@cs.washington.edu}

\begin{document}

\begin{abstract}
% \begin{ready-for-review}

% end-to-end hardware and software

We present ClearBuds, the first  hardware and software  system that utilizes a neural network to enhance speech streamed from two wireless earbuds. Real-time speech enhancement for wireless earbuds  requires high-quality sound separation and background  cancellation, operating in real-time and on a mobile phone.  ClearBuds bridges state-of-the-art deep learning for blind audio source separation and in-ear mobile systems by making two key technical  contributions: 1) a new wireless earbud design capable of operating as a synchronized, binaural microphone array, and 2) a lightweight dual-channel speech enhancement neural network that runs on a mobile device. {Our neural network has a novel cascaded architecture that combines a time-domain conventional neural network with a spectrogram-based frequency masking neural network to reduce the artifacts in the audio output.}  Results show that our wireless earbuds  achieve a  synchronization error less than 64 $\mu$s and 
our network has a runtime of 21.4 ms on an accompanying mobile phone. In-the-wild evaluation with eight users in  previously unseen indoor and outdoor multipath scenarios demonstrates that our neural network generalizes to learn both spatial and acoustic cues to  perform noise suppression and background speech removal. In a  user-study with 37 participants  who spent over 15.4  hours rating  1041   audio samples collected in-the-wild, our system achieves improved mean opinion score and background   noise  suppression.

\begin{center}Project page with demos: {\textcolor{blue}{{{\url{https://clearbuds.cs.washington.edu/}}}}}\end{center}

%The link below shows a system demo:\begin{center}
%\textcolor{blue}{{{\url{https://youtu.be/0Hmnc054cow}}}}
%\end{center}

\end{abstract}

\acmYear{2022}\copyrightyear{2022}
\acmConference[MobiSys '22]{The 20th Annual International Conference on Mobile Systems, Applications and Services}{June 25--July 1, 2022}{Portland, OR, USA}
\acmBooktitle{The 20th Annual International Conference on Mobile Systems, Applications and Services (MobiSys '22), June 25--July 1, 2022, Portland, OR, USA}
\acmPrice{15.00}
\acmDOI{10.1145/3498361.3538933}
\acmISBN{978-1-4503-9185-6/22/06}

 %  real-time mobile deep learning, binaural earbuds}



\maketitle

\section{Introduction}

\begin{figure}[t!]
\centering
 \includegraphics[width=1\linewidth]{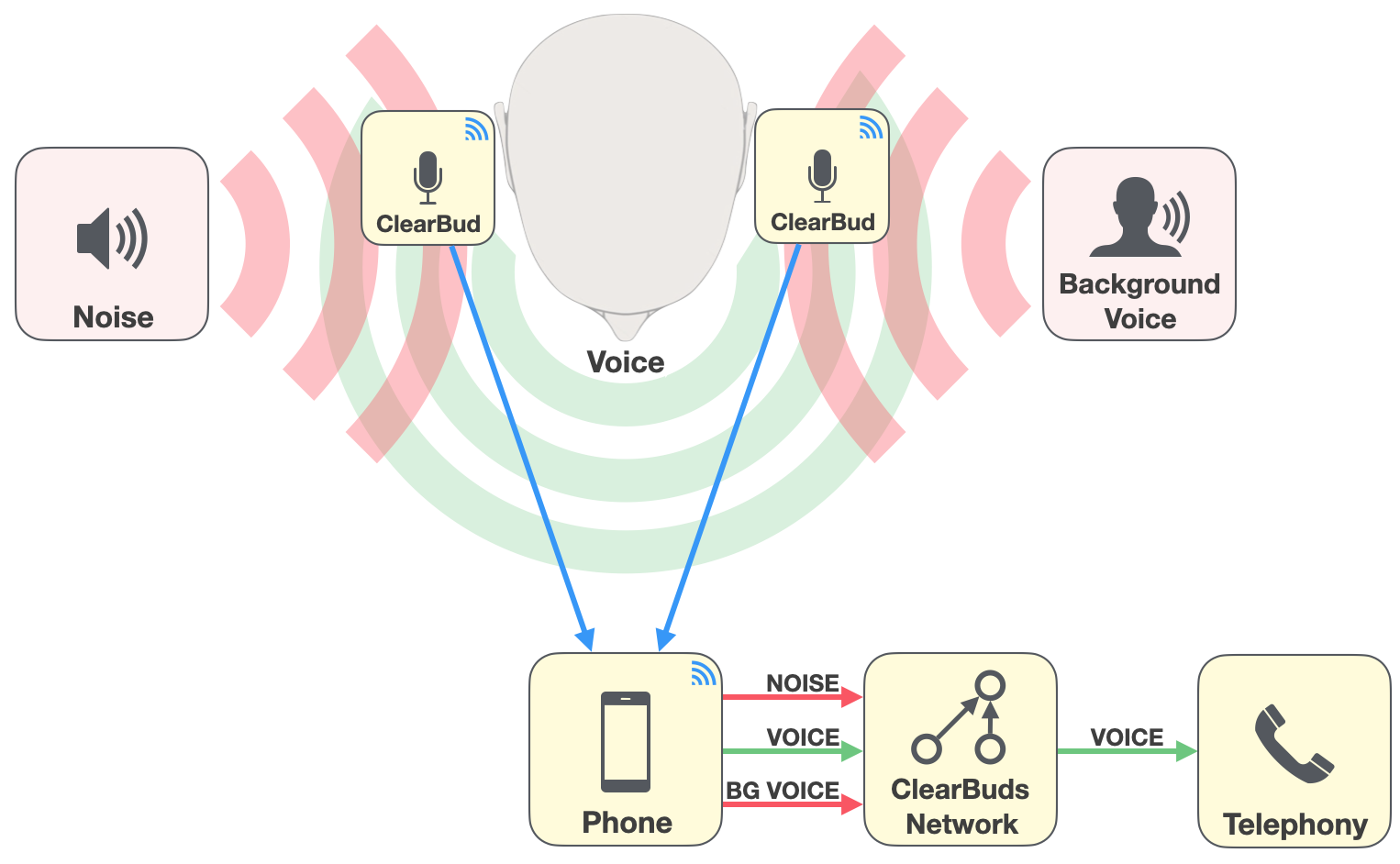}
\vskip -0.1in
\caption{{ ClearBuds Application.  Our goal is to isolate a user's voice from background noise (e.g., street sounds or other people talking) by performing source separation  using a pair of custom designed, synchronized, wireless earbuds.}}
\label{fig:block_diagram}
\vskip -0.15in
\end{figure}

% From a user experience perspective, this would allow a user to not toggle the mute button on call, even when they are not speaking.

%%%%%%%%%%%%%%%%%%%%%%%%%%%%%% DRAFT 4 %%%%%%%%%%%%%%%%%%%%%%%%%%%%%%%%%%%%

With the rapid proliferation of wireless earbuds (100 million AirPods sold in 2020~\cite{airpodssales}), more people than ever are taking calls on-the-go.  While these systems offer unprecedented convenience, their mobility raises an important technical challenge:  environmental noise (e.g., street sounds, people talking)  can interfere and make it harder to understand the speaker.
We therefore seek to enhance the speaker's voice and suppress background sounds using speech captured across the two earbuds.

Source separation of acoustic signals  is a long-standing problem where the conventional approach for decades has been to perform beamforming using multiple microphones. Signal processing-based beamformers that are computationally lightweight can encode the spatial information  but do not effectively capture acoustic cues~\cite{van1988beamforming,krim1996two,chhetri2018multichannel}.  Recent work has shown that deep neural networks can encode both spatial and acoustic information and hence can  achieve superior source separation  with gains of up to $9$~dB over signal processing baselines~\cite{subakan2021attention,luo2019conv}. However, these neural networks are computationally expensive. None of the existing binaural (i.e., using two microphones) neural networks can  meet the end-to-end latency required for telephony applications or have been evaluated with real earbud data. Commercial end-to-end systems, like  Krisp~\cite{krisp}, use neural networks on a cloud server for single-channel  speech enhancement, with implications to  cost and privacy. % or on edge devices.

%\footnote{The task of enhancing the wearer's  voice is for  telephony and video conferencing applications. This is not to be confused with active noise cancellation (ANC) which cancels out  noise coming to one's ears from an external environment~\cite{mute}.}

We present the first mobile system that uses neural networks to achieve real-time speech enhancement from binaural wireless earbuds.
Our key insight is to treat wireless earbuds as a binaural microphone array,
and exploit the specific geometry -- two well-separated microphones behind a proximal source -- to devise a specialized neural network for high quality speaker separation. In contrast to using multiple microphones on the same earbud to perform beamforming, as is common in Apple AirPods \cite{airpods} and other hearing aids, we use microphones across the left and right earbuds, increasing the distance between the two microphones and thus the spatial resolution.

{To realize this vision, we need to address three key technical challenges to deliver a functioning, practical system:
\begin{enumerate}
    \item Today's wireless earbuds only support one channel of microphone up-link to the phone. AirPods and similar devices upload microphone output from only a single earbud at a time.  To achieve binaural speaker separation, we need to design and build novel earbud hardware that can synchronously transmit audio data from both the earbuds, and maintain tight synchronization over long periods of time.
    \item Binaural speech enhancement networks have high computational requirements, and have not been demonstrated on mobile devices with data from wireless earbuds. Reducing the network size naively often leads to unpleasant artifacts. Thus, we also need to optimize the neural networks to run in real-time on smart devices that have a limited computational capability compared to cloud GPUs. Further, we need to meet the end-to-end latency requirements for telephony applications and ensure that the resulting audio output has a high quality from a user experience perspective.
    \item Prior binaural speech enhancement networks are trained and tested on synthetic data and have not been shown to generalize  to real data. Building an end-to-end system however requires a network that generalizes to in-the-wild use. 
\end{enumerate}}

%Achieving this goal is challenging for three key reasons. First, today's earbuds are not capable of operating in this manner; AirPods and similar devices upload microphone output from only a single earbud at a time.  To achieve binaural speaker separation, we  need to design and build novel earbud hardware that can synchronously transmit audio data from both the earbuds. Second, binaural speech enhancement networks are not lightweight and have not been demonstrated with wireless earbuds. Reducing the network size naively often leads to unpleasant artifacts. Thus, we also need to optimize the neural networks to run in real-time on smart devices that have a limited computational capability compared to cloud GPUs. Further, we need to  meet the end-to-end latency requirements for telephony applications and ensure that the resulting audio output has a high quality from a user experience perspective. {Third, prior binaural  speech enhancement networks are trained and tested  on synthetic data and have not been shown to generalize well to real  data. Building an end-to-end system however requires a network that generalizes to in-the-wild real use.} 

To achieve this system, we make three  technical contributions spanning earable hardware and neural networks.

\begin{figure}[t!]
\vskip -0.1in
\centering
\includegraphics[width=1\linewidth]{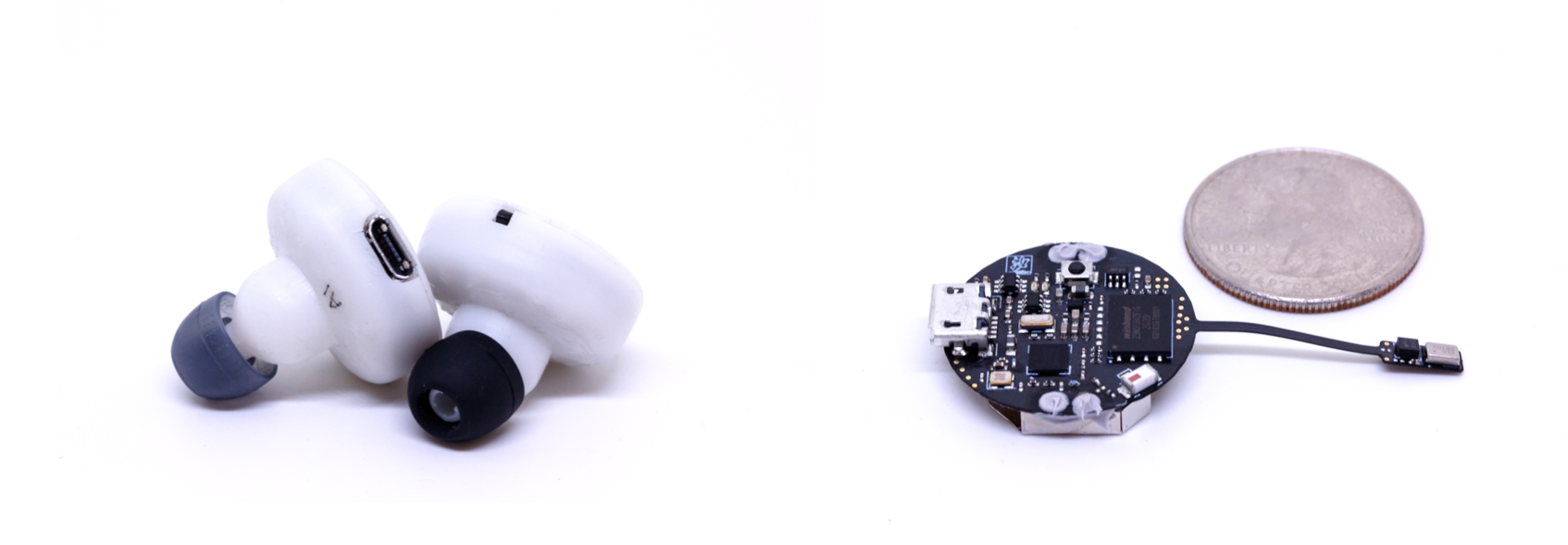}
\vskip -0.15in
\caption{{ClearBuds hardware  inside 3D-printed enclosure and when  placed beside a quarter. }}
\label{fig:earbuds}
\vskip -0.15in
\end{figure}

\squishlist
\item {\bf Synchronized  binaural earables.} We designed a binaural wireless earbud system  (Fig.~\ref{fig:earbuds}) 
capable of streaming two time-synchronized microphone audio streams to a mobile device. This is one of the first systems of its kind, and we expect our  {open-source} earbud hardware and firmware  to be of wider interest as a research and development platform. Existing earable platforms such as eSense~\cite{esense-1}  do not support time-synchronized audio transmission from two earbuds to a mobile device. 
We designed our DIY hardware  using open source eCAD software, outsourced fabrication and assembly ($\$2$K for 50 units), and 3D printed the  enclosures.

\begin{figure}[t]
\vskip -0.1in
\centering
\includegraphics[width=1\linewidth]{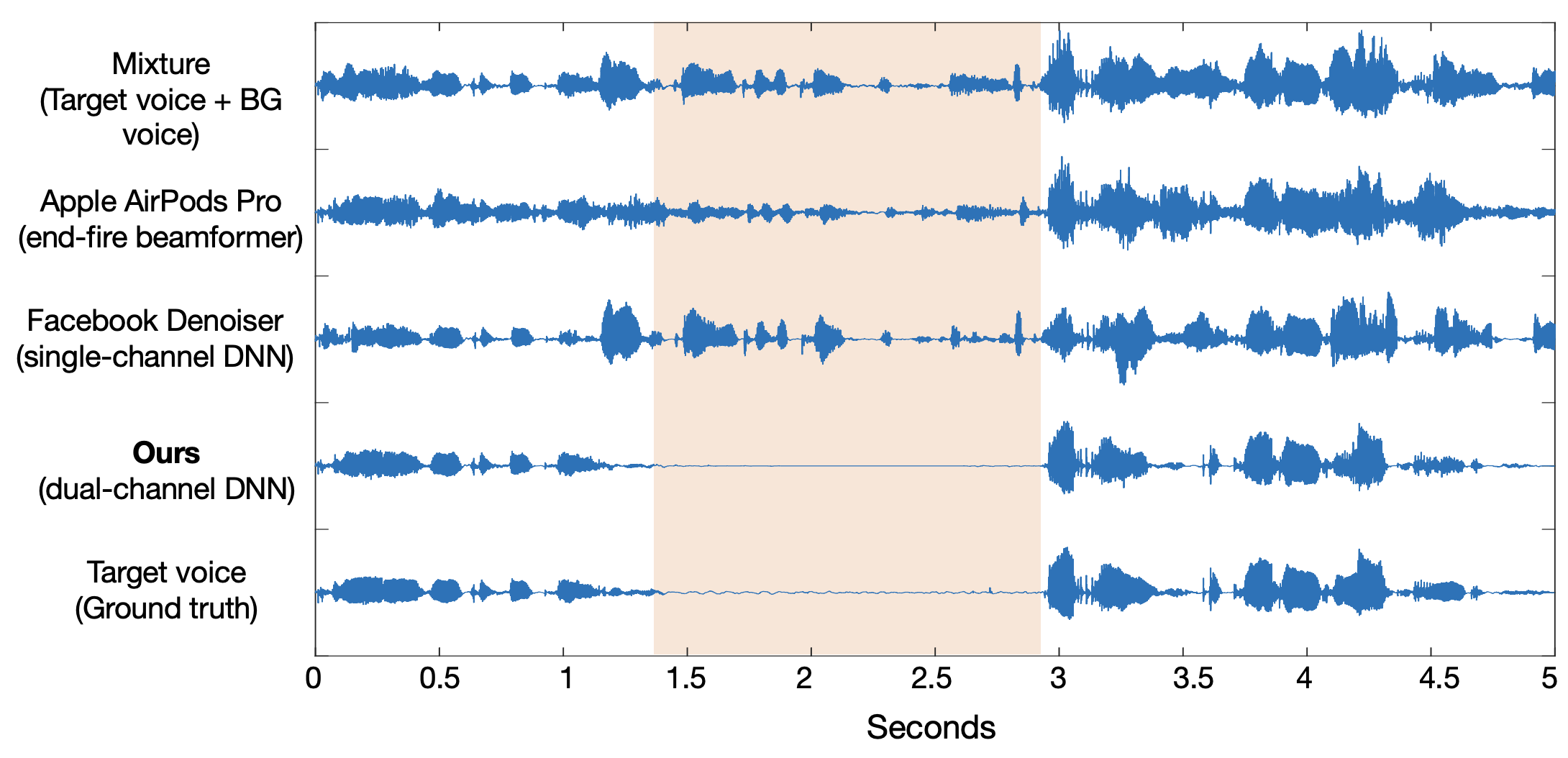}
\vskip -0.15in
\caption{{Background voice performance. { We use spatial cues to  separate background voices from the target speaker, even when the background voice is louder than the target voice. This is evident when the target speaker is silent  but background voice continues to talk (highlighted in orange).  Apple AirPods Pro  uses an endfire beamformer to partially suppress  background voice. The mono-channel Facebook Denoiser (Demucs) is unable to suppress the background voice. Clearbud's network  removes the background voice, approaching  ground truth. }}}%Comparing with Airpods:  \textcolor{blue}{\url{https://youtu.be/ZNhbs6lj6a0}} }}} %updated link 2022-05-25 11:48
\label{fig:fig3}
\vskip -0.2in
\end{figure}

\item {\bf Lightweight cascaded neural network.} We introduce a lightweight  neural network  that utilizes binaural input from wearable earbuds to isolate the target speaker. 
To achieve real-time operation, we start with the Conv-TasNet source separation network~\cite{luo2019conv} and redesign the network to achieve a 90\% re-use of the computed network activations from the previous time step for each new audio segment (see~\xref{sec:nn}). While these optimizations make this network real-time, they also introduce artifacts in the audio output {(i.e., crackling, static). Interestingly, these artifacts have little effect on traditional metrics, like Signal-to-Distortion Ratio (SDR), but have a noticeable effect on subjective listening scores (see \xref{sec:mos}). These artifacts however are often visible in a frequency representation of the audio. } To address this, we combine our mobile temporal model with a real-time  spectrogram-based frequency masking neural network. We show that by combining the two networks and creating a lightweight cascaded  network,  we can reduce artifacts and improve the audio quality further. 
\item {{\bf Network training for in-the-wild generalization.} 
Training the network in a supervised way requires clean ground truth speech samples as training targets. This is difficult to obtain in fully natural settings since the ground truth speech is corrupted with background noise and voices. Training a network that generalizes to in-the-wild scenarios also requires the training data to mimic the dynamics of real speech as closely as possible. This includes reverb, voice resonance, and microphone response. Synthetically rendered spatial data is the easiest type of data to obtain, but most different from real recordings, while real speakers wearing the headset in an anechoic chamber provide the best ground-truth training targets, but are the most costly to obtain. Synthetic data can simulate various reverb and multi-path that are not captured in an anechoic chamber. {Our training methodology  uses large amounts of synthetic data simulated in software, small amounts of hardware data with  speakers embedded into a foam mannequin head and small amounts of data from human speakers wearing the earbuds in an anechoic chamber   (see~\xref{sec:datasets}) 
to create a neural network that generalizes to  users and multi-path environments not in the training data. }}

 \squishend

We combine our wireless earbuds and neural network to create ClearBuds, an end-to-end system capable of (1) source separation for the intended speaker in noisy environments, (2) attenuation and/or elimination of both background noises and external human voices, and (3) real-time, on-device processing on a commodity mobile phone paired to the  two earbuds.  Our results show that:
\squishlist
    \item Our binaural wireless earbuds  can stream audio to a phone with a synchronization error less than 64$\mu$s and operate  continuously on a coin cell battery for 40 hours.
    \item Our system outperforms Apple AirPods Pro by 5.23, 8.61, and 6.94~dB  for the tasks of  separating the target voice from background noise, background voices, and a combination of background noise and voices respectively.
    \item Our network has a runtime of 21.4ms on iPhone 12, and the entire ClearBuds system operates in real-time with an end-to-end  latency of 109ms.  For telephony applications, an ear-to-mouth latency of less than 200ms is required for a good user experience \cite{g.114}.
    \item In-the-wild evaluation with eight users in various indoor and outdoor scenarios  shows that our system  generalizes to previously unseen participants and multipath environments,  that are not in the training data.  
    \item In a user study with 37  participants who spent over 15.4  hours and rated a total of 1041  in-the-wild  audio samples, our cascaded network achieved a higher mean opinion score and noise suppression  than both the input speech as well as a  lightweight Conv-TasNet.

%    \item Our network can run in real-time on a mobile device with a latency of 155ms on an iPhone 12
\squishend

We believe that this paper  bridges state-of-the-art deep  learning for blind audio source separation and in-ear mobile systems. The ability to perform background noise suppression and   speech  separation could positively   impact  millions of people who use earbuds to take calls on-the-go. By open-sourcing the hardware  and collected datasets, our work may  help kickstart future research among mobile system and machine learning researchers to design  algorithms around wireless earbud data.

\section{Related Work}

%\begin{help-env}

Endfire beamforming configurations remain popular on  consumer mobile phones and earbuds \cite{samsungglobalnewsroom_2014, airpods, sennheiser_2020, beamforming-app-note}. While recent advances in neural networks have shown promising results, none of them are demonstrated with  wireless earbuds.  By creating a wireless   network between two earbuds, we demonstrate that our real-time, two-channel neural network can outperform  current real-time speech enhancement approaches for wireless earbuds. %Below, we briefly discuss beamforming, single channel speech enhancement, and binaural networks.

%A common approach to enhancing speech is to design a beamforming microphone array to be more sensitive to sounds coming from the direction of the user's mouth \cite{van1988beamforming} or voice~\cite{dov-uist21}.

\vskip 0.05in\noindent{\it Beamforming techniques.}  Since signal-processing based beamforming is computationally lightweight, these techniques are deployed on  commercial devices such as smart speakers \cite{amazon}, mobile phones \cite{samsungglobalnewsroom_2014}, and earbud devices like Apple AirPods \cite{airpods}. However, the performance of beamforming is limited by the geometry of the microphones and the distance between them \cite{van1988beamforming, InvenSense}. The form factor of devices like AirPods restricts both the number of microphones on a single earbud and the available distance between them, limiting the gain of the beamformer. While beamforming  across two earbuds could provide better performance in principle, current wireless architectures are limited to streaming from a single earbud at a time \cite{bluetooth}. Furthermore, adaptive beamformers such as MVDR \cite{frost1972MVDR}, while showing promise with relatively few interfering sources,  are sensitive to sensor placement tolerance and steering \cite{zhang2017deep, brandstein2001microphone}. Finally, beamforming leverages spatial or spectral cues only and does not use acoustic cues (e.g., structure in human speech) and perceptual differences to discriminate sources, information that machine learning methods leverage successfully.

% Don H. Johnson and Dan E. Dudgeon. Array Signal Processing: Concepts and Techniques. Simon & Schuster, Inc., USA, 1992

% Beamforming microphone arrays for speech enhancement, https://ieeexplore.ieee.org/abstract/document/225915

% Rate-Constrained Beamforming in Binaural Hearing Aids
% https://link.springer.com/content/pdf/10.1155/2009/257197.pdf

% Dual-Channel Speech Enhancement by
% Superdirective Beamforming
% https://link.springer.com/content/pdf/10.1155/ASP/2006/63297.pdf

%\vskip 0.05in\noindent{\bf Single-channel speech enhancement.} 
{\vskip 0.05in\noindent{\it Single-channel deep speech enhancement.}
Many deep learning techniques operate on spectrograms to separate the human voice from background noise \cite{realtimenoise, Mohammadiha_2013, online_nonnegative, nikzad2020deep, choi2019phaseaware, lstm_speechenhancement, fu2019metricgan, TFMasking}. However, recent works  instead operate directly on the time domain signals \cite{luo2019conv, germain2018speech, pascual2017segan, demucsreal, macartney2018improved}, yielding performance improvements over spectrogram approaches. Commercial noise suppression software like Krisp \cite{krisp} and Google Meet \cite{googlemeet} have successfully deployed single-channel models in real-time and are available for use on mobile phones and desktop computers, but {processing is performed on the cloud.}~\cite{tinylstm}  achieves low-power speech enhancement using
long short-term memory (LSTM), but it is for a  single-channel network but not for multichannel source separation. Further, single-channel models cannot effectively capture  spatial information and fail to isolate the intended speaker when there are multiple speakers (see Fig.~\ref{fig:fig3}).}

% --> some recent work on increasing performance to real-time applications
% Real Time Speech Enhancement in the Waveform Domain
% https://arxiv.org/abs/2006.12847

\begin{figure*}[t!]
\vskip -0.1in
\centering
\includegraphics[width=1\linewidth]{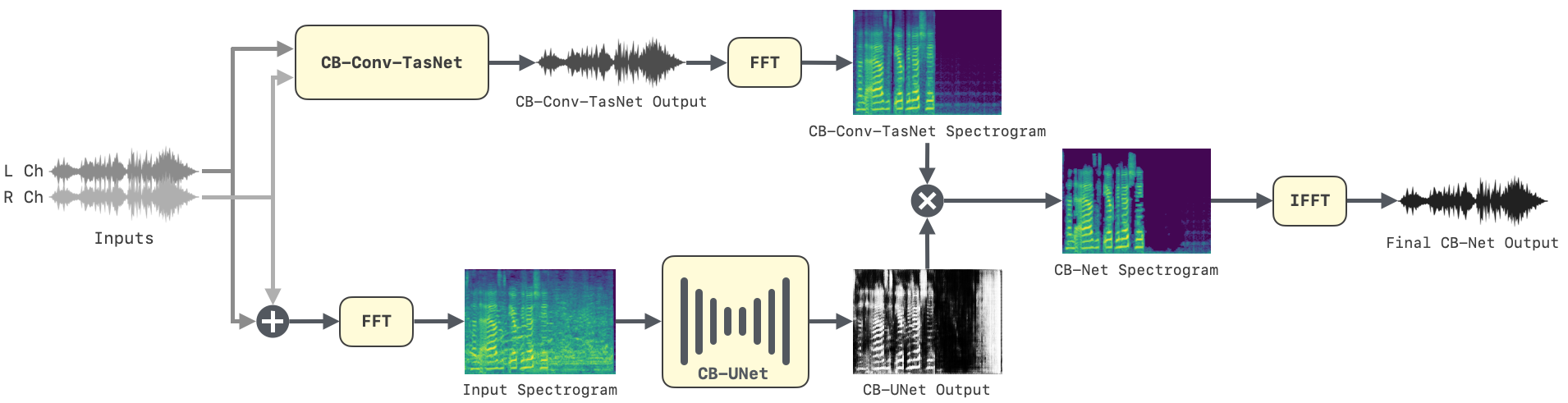}
\vskip -0.14in
\caption{Network Diagram of CB-Net. Our network contains a time-domain component, shown in the top as CB-Conv-TasNet, and a frequency domain component, shown on the bottom by CB-UNet. } %The respective outputs are combined to produce the final output.}
%\vskip -0.12in
\label{fig:network-full}
\end{figure*}

{\vskip 0.05in\noindent{\it Multi-channel source separation and speech enhancement.}
%\subsection{Multi-channel source separation and speech enhancement}
Multi-channel methods have been shown to perform better than their single-channel source separation counterparts \cite{yoshioka2018multi, chen2018multi, zhang2017deep, gu2020enhancing, tzirakis2021multichannel, jenrungrot2020cone}. Binaural methods have also been used for source separation \cite{binaural1, han2020realtime, li2011two, reindl2010speech} and localization \cite{van2008binaural, lyon1983computational, kock1950binaural}; \cite{han2020realtime} reduces the look-ahead time in the network to make it causal in behavior but has not been demonstrated to run on a mobile device.  Our method improves on existing binaural methods by combining time-domain neural network with spectrogram-based frequency masking networks as well as optimizing them to enable   real-time processing on a phone. Recent works such as ~\cite{binaural_osu, dual_phone,binauralphone} use multiple microphones on a smartphone for  speech-enhancement. However, neither of them demonstrates evaluation with real data, where artifacts because of network optimizations can affect user performance.
In contrast, we demonstrate the first system that achieves real-time speech enhancement using microphones on the two wireless earbuds. Further, as the distance between the earbuds is larger than the distance between microphones on a typical mobile phone, we can  attain a better baseline than a mobile phone implementation, while also retaining the ability to speak hands-free. More recent  works tackle the problem of real-time directional hearing using eye trackers and wearable headsets. For example,  \cite{hybridbeam} uses a hybrid network that combines signal processing with neural networks, but shows that their technique performs poorly in binaural scenarios (i.e., two microphones) and requires four or more microphones. In contrast, we focus on the problem of speech enhancement and create the first real-time end-to-end hardware-software neural-network based system using wireless synchronized earbuds. }

\vskip 0.05in\noindent{\it  Earbud  computing and platforms.}
There has been recent interest in earbud computing~\cite{oesense21,esense-1,esense-2,plat-1,romit-1} to address  applications in health monitoring~\cite{infection,tam1,tymp}, activity tracking~\cite{mobisys21} and sensor fusion with EEG signals~\cite{eeg2}. The  eSense platform~\cite{esense-1,esense-2} has enabled research in   sensing applications with earables. OpenMHA~\cite{open-1,open-2} is an open signal processing {\it software} platform for hearing aid research. Neither of these  platforms   support time-synchronized audio transmission from two earbuds, which is a critical requirement for achieving  speech enhancement in binaural settings.    In contrast, we created open-source wireless earbud hardware that can support synchronize wireless transmission from the two earbuds.

\section{ClearBuds Design}

We first introduce our  lightweight  neural network architecture. We then describe system design  of our hardware platform and our synchronization algorithm. % to ensure time alignment between each ClearBud's microphone.
We open-source our mechanical, firmware, application, and network designs at our project website: \textcolor{blue}{{{\url{https://clearbuds.cs.washington.edu}}}}.

\subsection{Problem Formulation}
Suppose we have a 2 channel microphone array with one microphone on each ear of the wearer. The target voice is speaking with a signal $s_0 \in \mathbb{R}^{2 \times T}$ in the presence of some background noise $\textbf{bg}$ or other non-target speakers $s_{1..N}$. There may also be multi-path reflections and reverberations $\textbf{r}$ which we would also like to reduce, i.e., $\mathbf{x} = \sum_{i=0}^{N}\mathbf{s}_i + \mathbf{bg} + \mathbf{r}$.  Our goal is then to recover the target speaker's signal, $s_{0}$, while ignoring the background, reverbations, or other speakers. We also must do so in a real-time way, meaning that the a mixture sample $\textbf{x}_t $ received at time $t$ must be processed and outputted by the network before $t + \textbf{L}$ for some defined latency $\textbf{L}$.  We refer to the non-target speakers as "background voices". These background voices may be at any location in the scene, including very close to the target speaker and their angle  can change with time and motion.

\subsection{Neural Network Architecture Motivation}\label{sec:nn}

 %Our target network is approximately causal and runs in real-time on a mobile device.
  {Our network needs to perform in real-time on a mobile device with minimal latency.}
 This is challenging for several reasons. First, the processing device has a much lower compute capacity, especially compared to cloud GPUs. Additionally, the network should separate non-speech noises as well as unwanted speech. To do this, it must learn spatial cues and human voice characteristics. Finally, the resulting output should maximize the quality from a human experience perspective while minimizing any artifacts the network might introduce. 

Our network, which we call \textit{ClearBuds-Net} or \textit{CB-Net}, is a cascaded model that operates in both  time  and frequency domains. The full network architecture is illustrated in Fig.~\ref{fig:network-full} and contains two main sub-components: A dual-channel time domain network called \textit{CB-Conv-TasNet}, and a frequency based network called \textit{CB-UNet}. %Next, we  describe the motivation for each component of the network.

% \textcolor{red}{XXXThis needs to be rewritten. We need tos tart with the network architecutre and then talk about the details of each of the network and show example spectrograms to give the motivation for the frequency masking network.}

% To address this,. Second, the compute device has a much lower memory and compute capacity, especially compared to standard GPUs. To solve this, we use depthwise separable convolutions as described in \cite{howard2017mobilenets}. We also use a temporal convolution network (TCN) that allows caching of intermediate outputs to reduce computation on new packets. Finally, 

\subsubsection{CB-Conv-TasNet}
The first component of separation method is a time domain network that is based on a multi-channel extension of Conv-TasNet \cite{luo2019conv}. This is a network in the waveform domain that  has a  Temporal Convolution Network (TCN) structure, lending itself to a causal implementation with intermediate layer caching \cite{paine2016fast}. We  use depthwise separable convolutions~\cite{howard2017mobilenets} to  further reduce the number of parameters and make the design real-time. We call this network CB-Conv-TasNet since it is an optimized version of the original Conv-TasNet.

A key feature of the time domain approach is that it can easily capture spatial cues in the network. In our application, the desired source is always physically between  two  microphones, thus the voice signal will reach the microphones roughly at the same time. In contrast,  background or other speakers are typically not temporally aligned and will reach one microphone earlier or later. By feeding two time synchronized channels into the neural network, this spatial alignment of the sources can be learned from time differences in the signal. This is similar to a delay-and-sum beamforming effect, except the sum is replaced with a deep network. %Although a similar spatial separation approach was demonstrated in \cite{jenrungrot2020cone}, this spatial approach in the time-domain alone is not sufficient to produce quality output given the resource constraints of the computing device. NOT SURE WHAT THIS IS ADDING EXCEPT FOR HIGHLIGHT THIS PRIOR WORK.

\begin{figure*}[t!]
%\vskip -0.5in
\centering
\includegraphics[width=1\linewidth]{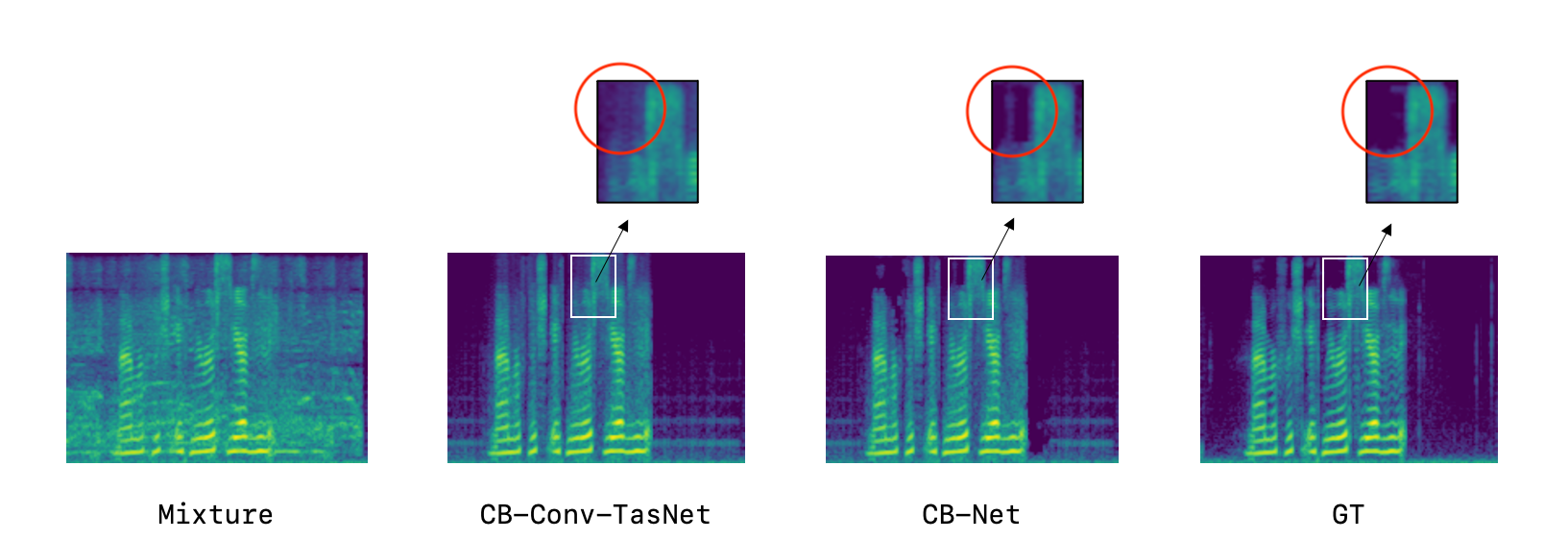}
\vskip -0.15in
\caption{{The spectrograms above show the motivation behind a combined time and frequency domain method. The output of the time-domain component, CB-Conv-TasNet, contains artifacts, particularly at high frequencies. Although subtle, these artifacts are perceptible by human listeners. CB-Net is able to reduce these artifacts by using a frequency-domain network (CB-UNet) that masks unwanted frequencies.}}
\vskip -0.2in
\label{fig:spectrogram-motivation}
\end{figure*}

\subsubsection{CB-UNet}
 {The output of our lightweight CB-Conv-TasNet often contains audible artifacts (i.e., crackling, static) that reduce the listening experience. Interestingly, these artifacts have little effect on traditional metrics, like Signal-to-Distortion Ratio (SDR), but have a noticeable effect on subjective listening scores (see \xref{sec:mos}). These artifacts are often visible in a frequency representation of the audio.  Fig.~\ref{fig:spectrogram-motivation} shows how  CB-Conv-TasNet alone contains noticeable artifacts when compared to the ground truth. To address this, we cascade a lightweight causal UNet \cite{ronneberger2015unet} which operates on the mel-scale spectrogram of the input audio. This network, which we call CB-UNet, produces a binary mask which is applied to the output of CB-Conv-TasNet. The combined output, shown in Fig.~\ref{fig:spectrogram-motivation} as CB-Net,  reduces these artifacts. The mean opinion scores in our evaluation  shows the strength of the cascaded CB-Net when compared to the time-domain component only.}

\subsection{Neural Network Detailed Description}

%Here, we describe details of our  network architecture.

\subsubsection{CB-Conv-TasNet}
The input to the network is a binaural mixture given by $\textbf{x} \in \mathbb{R}^{2 \times T}$. The first step is an encoder that transforms the mixture $\textbf{x}$ into $\mathbb{R}^{N \times T/L}$ with a 1D convolution of size $L$ and stride $L$. This is followed by a ReLU layer. The encoder's outputs are next fed into a temporal convolution network that consists of stacks of 1-D convolutions with increasing dilation factors. We use 14 convolution layers with dilation factors of 1,2,4,..,64 repeated twice, with a ReLU nonlinearity and skip connection after each convolution. The encoder output is multiplied with the output of the temporal conv-net, before being fed through a fully connected Decoder layer which transforms the output back into $\mathbb{R}^{2 \times W}$. %We describe window size  $W$ in the next sections. %\textcolor{red}{You can also refer to figure (need to insert) for a visual layout of the network.}

In a real-world implementation, we do not have access to the full waveform, but only packets of data at a time. Furthermore, we must process these packets with limited access to future input samples. Given  15.625 kHz sampling rate, we choose to process packets of 350 samples at a time (22.4ms), which is our  window size $W$. We also use $2W$, or 700 samples of lookahead time (44.8ms) and 1.5s of past samples. Since we have no padding in the temporal convolution net, the network starts with this large temporal context and outputs exactly $\mathbb{R}^{1 \times W}$ samples, corresponding to the desired output for our input packet of $W$ samples. When we receive the next packet of size $W$, all intermediate activation from the encoder and temporal conv-net can be shifted over by $W / L$ samples and re-used. We chose $L = 50$, but any divisor of $W$ would work. Re-using intermediate outputs from previous packets saves over $90\%$ of the compute time for a new packet in our network. %These ideas of caching causal convolutions has been discussed in previous work such as \cite{paine2016fast}. %https://arxiv.org/pdf/1611.09482.pdf%

\begin{figure}[t!]
%\vskip -0.1in
\centering
\includegraphics[width=1\linewidth]{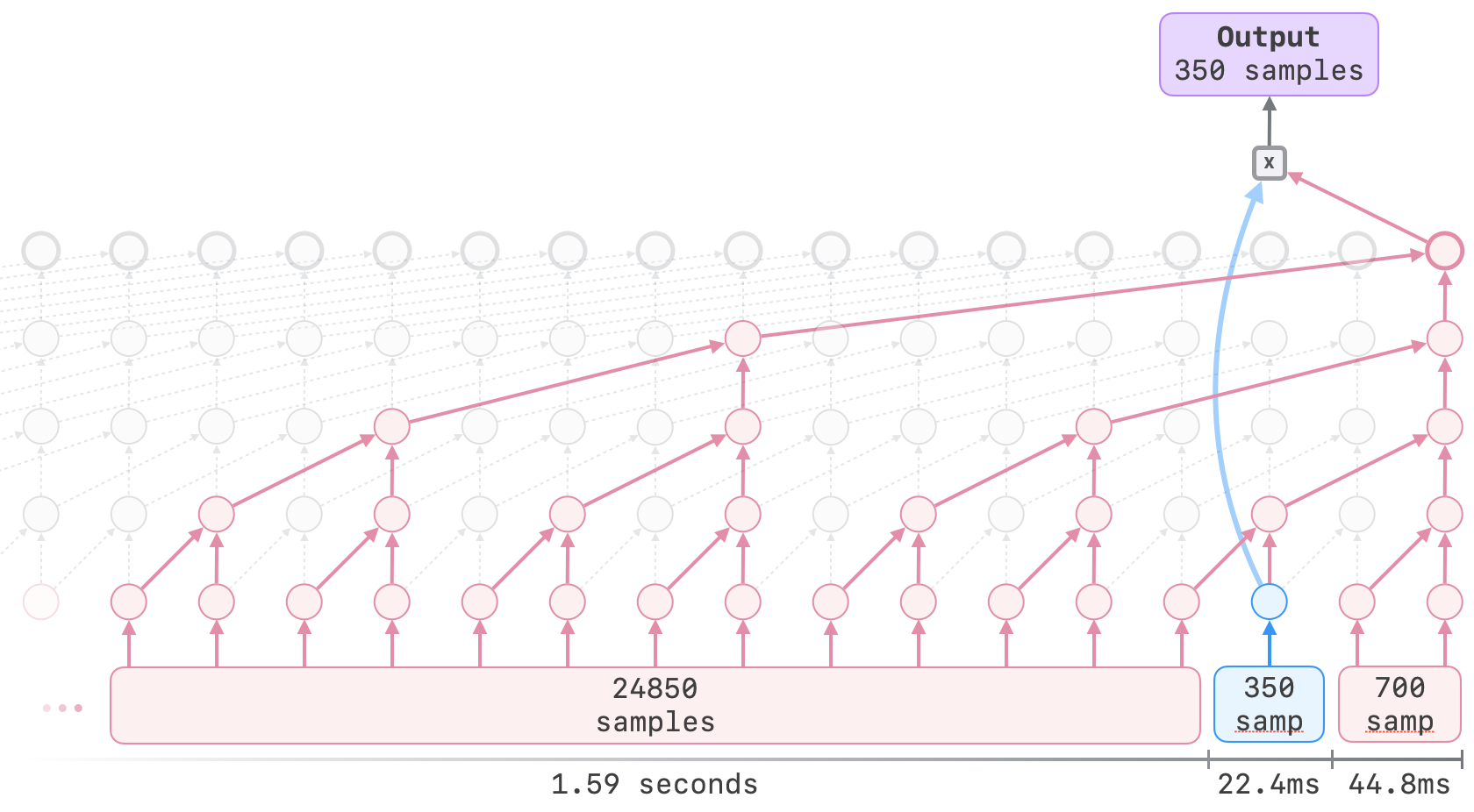}
\vskip -0.15in
\caption{ CB-Conv-TasNet, the time-domain component of CB-Net. Given a packet of 350 samples (22.4ms) highlighted in blue, we use 1.5s of past input and 44.8ms of future input to output the separation results. Our caching scheme works as follows: When we receive a new 350ms samples, all intermediate activations (circles in the diagram) slide to the left, and we compute only the rightmost column of outputs.}
\label{fig:network}
\vskip -0.3in
\end{figure}

\subsubsection{CB-UNet}

The frequency domain network is a mono-channel network that outputs a binary mask for each time-frequency bin. The input $\textbf{x} \in \mathbb{R}^{1 \times T}$ is a summation of the binaural left and right channel, which is the equivalent of a broadside beamformer. We first run a STFT, which is a mel-scale fourier transform with hop size of $350$, a window size of  $1024$ including zero padding on the edges, and a $128$ bin mel-scale output. The network input is a spectrogram of $64$ time bins and $128$ frequency bins, corresponding to a receptive field of $22400$ samples, or $1.43s$. In order to maintain the causality requirement, we use the same lookahead strategy as the time-domain network where we allow $700$ samples of lookahead for a target packet of $350$ samples. The UNet architecture contains 4 downsampling and upsampling layers, starting with 64 channels and doubling the number of channels at each subsequent layer. The downsampling layers contain a depthwise separable convolution followed by a $2 \times 2$ max pooling, and the upsampling layers contain a depthwise separable convolution followed by a transposed convolution for upsampling. The output is a sigmoid function, which is then thresholded to return a binary mask in $[0, 1]^{128 \times 64}$.  When outputting a spectrogram mask on an $\mathbb{R}^{128 \times 64}$ input, we predict a mask over the entire input even though we only need the output for a specific slice of $350$ samples, or a $\mathbb{R}^{128 \times 1}$ mask. Further optimizations could be made by caching intermediate outputs or only computing the mask for the target samples. However CB-UNet's run-time  was so small compared to the rest of the network that these optimizations were not  necessary.

\subsubsection{Combining the Outputs}
At each time step, the output of CB-Conv-Tasnet is an audio waveform in $x \in \mathbb{R}^{1 \times 350}$, and the output of CB-UNet is a spectrogram mask in $\textbf{M} \in \mathbb{R}^{128 \times 64}$. We run the same fourier transform on the buffered conv-tasnet outputs to produce a spectrogram $\textbf{X} \in \mathbb{R}^{128 \times 64}$. Our output can then be computed by $iSTFT(\textbf{M} \otimes \textbf{X})$.
 Our empirical results show that  this gives the best results compared to other methods such as ratio masking.
%Previous work \cite{zhao2018sound} has shown the benefits of binary masking for spectrogram based separation, and

\subsubsection{Training}
CB-Conv-TasNet is trained with an $L1$-based loss over the waveform along with the multi-resolution spectrogram loss. Formally, provided $s_{0}$ is our target speaker and $x'$ is the output from the network, our loss is:
\begin{equation*}
    L(s_{0}, x') = \left\Vert s_{0} - x' \right\Vert_1 \\ + L_{sc}(s_0, x') + L_{mag}(s_0, x')
\end{equation*}
\begin{equation*}
L_{sc}(s_0, x') = \frac{\left\Vert STFT(s_0) - STFT(x')\right\Vert_F}{\left\Vert STFT(s_0)\right\Vert_F}
\end{equation*}
\begin{equation*}
L_{mag}(s_0, x') = \left\Vert \log(STFT(s_0) - \log(STFT(x'))  \right\Vert_1
\end{equation*}
STFT denotes the magnitude of the short time Fourier transform, and $F$ denotes the Frobenius norm. $L_{sc}$ and $L_{mag}$ represent  spectral convergence and magnitude losses, which  gave better results than L1 loss alone.

For training CB-UNet, for each time frequency bin, the training target $\textbf{M}$ is 1 if the target voice is the dominant component, and 0 otherwise. Formally, $\textbf{M}(f, t) = [\textbf{S}_{0}(f, t) \geq \textbf{S}_{i}(f, t)], \;\; \forall i = (1..n)$. The network is then trained with the binary cross entropy of the output compared to the target mask.

\subsubsection{Hyperparameters and Training Details}

 We use a learning rate of $\SI{3e-4}{}$ along with the ADAM optimizer  \cite{kingma2014adam} for training the network. The network was trained on a single Nvidia TITAN Xp GPU. Because of the small size of the network, training could be completed within a single day and generally required $\approx$50 epochs to reach convergence.As an additional data augmentation step we make the following perturbations to the data: High-shelf and low-shelf gain of up to $\SI{2}~{dB}$ are randomly added using the \texttt{sox} library.

\subsection{Synchronized wireless earbuds}

\begin{figure}[t!]
\vskip -0.1in
\centering
\includegraphics[width=1\linewidth]{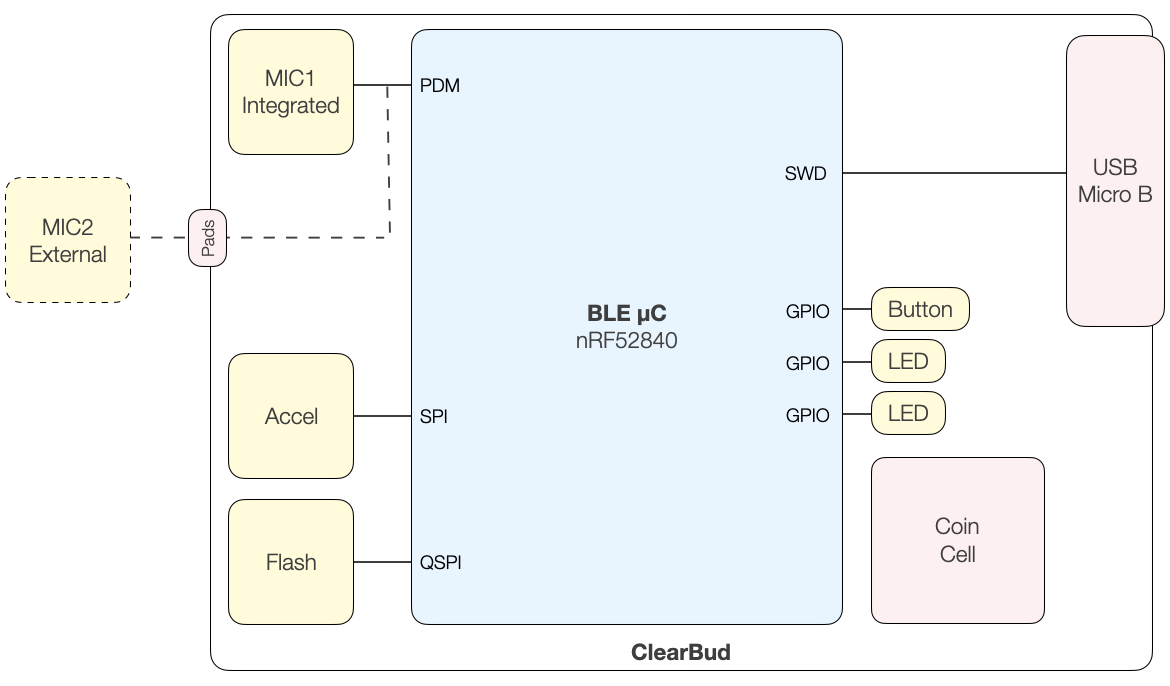}
\caption{Hardware Block Diagram. Each ClearBud integrates a PDM microphone, accelerometer, flash, and coin cell battery. Buttons and LEDs are used for interfacing with the device, and a USB port is used for programming and debug.}
\vskip -0.2in
\label{fig:block-diagram}
\end{figure}

\label{sec:system}
 We seek to capture speech from the target speaker's mouth which sits on the sagittal plane roughly  equidistant to the ears. Given an ear-to-ear spacing of 17.5cm, to effectively isolate this central plane we require a distance precision on the order of a few centimeters.
An interaural time difference of 100$\mu$s  would correspond to source maximally 3.43 cm off this central plane, therefore we target a synchronization accuracy under 100$\mu$s.

\begin{figure*}[t!]
\vskip -0.1in
\centering
\includegraphics[width=1\linewidth]{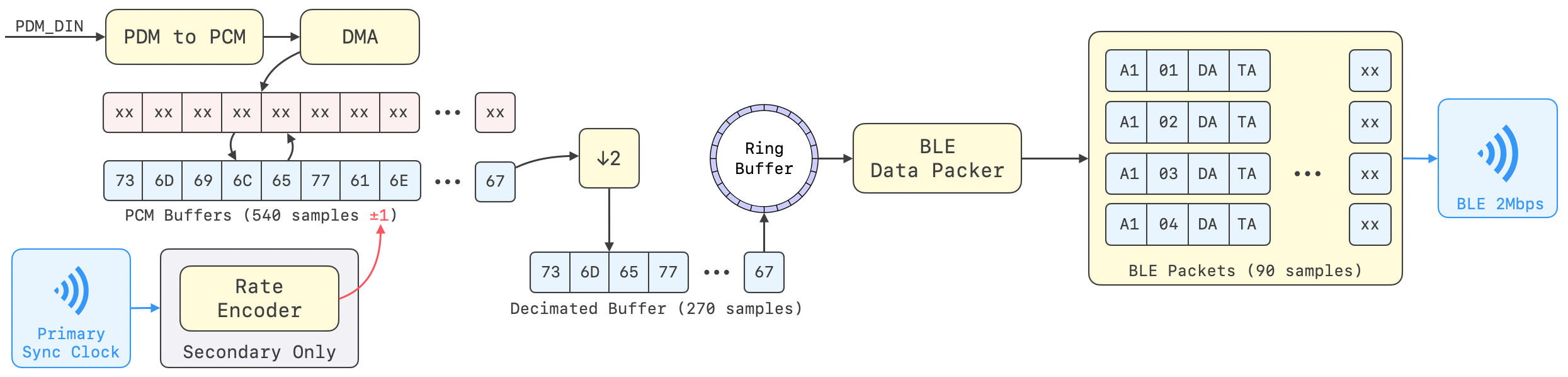}
\vskip -0.1in
\caption{Time Sync Design. The primary ClearBud broadcasts a sync clock over the air. The secondary ClearBud then uses the sync clock to rate encode, by increasing or decreasing the size of its local PCM buffer.}% Both ClearBuds then decimate these buffers down by a factor of 2, and fill up a ring buffer.} %Finally, each BLE packet gets prepended with a sequence number and then gets transmitted to a host device. }
\label{fig:sync}
\vskip -0.15in
\end{figure*}

\subsubsection{Hardware}  Our  custom hardware design contains  a %support for up to two 
pulse-density modulated (PDM) microphone (Invensense ICS-41350) and a Bluetooth Low Energy (BLE) microcontroller (Nordic nRF52840).\footnote{For future research applications, an ultra low-power accelerometer (Bosch BMA400), a 1Gbit NAND flash for local data collection (Winbond W25N01GVZEIG), and support for speaker and an additional microphone are included.} The system is powered off of a CR2032 coin cell battery and programmed via SWD over a Micro-USB connector.  Each ClearBud has an integrated PDM microphone set to a clock frequency of 2MHz. With an internal PDM decimation ratio of 64, this provides us a sampling frequency of 31.25kHz. As most HD voice applications and wideband codecs are limited to 16kHz~\cite{voicecodecs}, we decimate further in firmware by a factor of 2, giving us a final sampling frequency of 15.625kHz.

Two 16-bit 180 sample size Pulse-Code Modulation (PCM) buffers are round-robined: one is filled with incoming PCM data while the other is processed. The DMA is responsible for both clocking in the PDM data and converting it into PCM. One buffer is always connected to the DMA, while the other is freed for processing for the rest of the data pipeline. When the buffer connected to the DMA fills, the buffers switch roles and we begin processing data on the newly freed buffer, and connect the other buffer back to the DMA. With this design we always have a continuous PCM stream to operate on.
Both ClearBuds transmit the PCM microphone data to a mobile phone for input into our neural network. To maximize throughput, we use the highest Bluetooth rate and packet sizes supported by iOS, which is 2Mbps and 182 bytes, respectively. We design a lightweight wireless protocol where the first 2 bytes represent a monotonically-increasing sequence number, while the other 180 bytes are reserved for the 16-bit PCM audio samples. The sequence number is used on the  phone so that we can zero-pad PCM data in the occasional event that a packet is dropped either over-the-air or by the radio hardware. This zero-padding keeps the left and right microphone data aligned on the host side in areas of poor radio performance or interference in the environment.

%\subsubsection{Fabrication} 
The hardware schematic and layout for ClearBuds was designed using the open source eCAD tool KiCad. A 2-layer flexible printed circuit was  fabricated and assembled by PCBWay. The 3D printed enclosures were designed using AutoDesk Fusion 360 and printed with a Phrozen Sonic Mini   using  a liquid resin fabrication process.   The MEMS microphone sits behind the lid on the earbud's outer surface.
A single button on the enclosure  provides access to turn on and off the earbuds.

% The sequence number is used on the mobile phone so that we can zero-pad PCM data in the occasional event that a packet is dropped either over-the-air or at the radio hardware layer. This zero-padding keeps the left and right microphone streams aligned on the host side in areas of poor radio performance or interference. However, this is not sufficient to maintain synchronization within 100$\mu$s. We must also account for clock drift between our earbuds. MK: This sounds worse, as if this is the foundation for the sync and clock drift is the secondary feature.

%%Our system architecture revolves around the design of two microphones placed in each user's ear. We choose this design for two reasons. First, it allows us to leverage the spatial data of our desired voice being centered between the mics, while background noise has a higher likelihood of arriving off-center. Second, this configuration allows us to be directly applicable to devices in the hearable sector.

%One hurdle that needs to be overcome, is synchronizing the audio input from both wireless microphones so that our network can properly leverage the TDOA between the audio streams.

%  https://devzone.nordicsemi.com/nordic/short-range-guides/b/bluetooth-low-energy/posts/wireless-timer-synchronization-among-nrf5-devices

\subsubsection{Microphone synchronization} Three components are necessary for maintaining microphone synchronization: (1) As each of our earbuds has its own local clock source, we need to establish a common clock between them so that they have the same reference of time, (2) a synchronized startup so each earbud starts recording from their respective microphone at the exact same time, and (3) a rate encoding scheme to control the earbud's sampling rate to match each other. 

In our system, each earbud has its own respective 32MHz clock source with a total +/- 20ppm frequency tolerance budget. So, in the worst case scenario, the earbuds will have 2.4 milliseconds of drift each minute. We  use the  Nordic's TimeSlot API \cite{timeslot}, which grants us access to the underlying radio hardware in between Bluetooth transmissions. This provides us a transport to transmit and receive accurate time sync beacons \cite{wireless-timesync}. Each ClearBud keeps a free-running 16MHz hardware timer with a max value of 800,000, overflowing and wrapping around at a rate of about 20~Hz. One ClearBud is assigned as the timing master while the other ClearBud will synchronize its free-running timer to the master's.  The primary ClearBud (timing master) transmits time sync packets at a rate of 200~Hz. These packets contain the value of the free-running timer at the time of the radio packet transmission. When the secondary ClearBud receives this packet, it can then add or subtract an offset to its own free-running timer for a common clock. %This allows ClearBuds to maintain an accurate common clock. 

Once each ClearBud is connected to the mobile phone, the phone sends a \verb|START| command to both ClearBuds over BLE. Each ClearBud contains firmware which arms a programmable peripheral interconnect (PPI) to launch the PDM bus once the 16MHz free-running timer wraps around at 800,000. By using this method, we bypass the CPU and trigger a synchronized startup entirely at the hardware layer. One caveat is that the mobile phone could write to one ClearBud right \textit{before} its clock wraps around at 800,000, and the other ClearBud right \textit{after} it wraps around at 800,000. With a clock that wraps around at 20Hz, this would trigger a mismatched startup and cause an alignment error of 50ms. To correct for this, each ClearBud reports its common clock timer value to the phone once it has received the \verb|START| command. The phone can then remove the first 781 audio samples (781 samples / 15.625kHz = 50ms) if one ClearBud started streaming 50ms before the other.

The final component to keeping the audio streams aligned is to create a rate encoding scheme between the ClearBuds. With the time sync beacons from the primary ClearBud, the other ClearBud now has both its local clock and the common clock (primary ClearBud's local clock). With these two clocks, the secondary ClearBud can identify how much faster or slower its PDM clock is running in relation to the primary ClearBud. We note that with a 2MHz PDM clock and a PDM decimation ratio of 64, each audio sample occupies 32~us. The non-primary ClearBud can then add or remove a sample to its PDM buffer every time the difference between the clocks exceeds a multiple of 32~us. By doing this, the secondary ClearBud ensures that its PDM buffer starts filling up at the exact same time as the primary ClearBud's PDM buffer, with a tolerance of 32 us. 
% \subsubsection{Hardware evaluation} \textcolor{red}{This should go to the evaluation. Have to talk about this...} In order to validate our synchronization firmware, we place both ClearBuds roughly equidistant from a speaker. A click tone is played every 15 seconds for 5 minutes, and recorded on both ClearBuds with time sync disabled and enabled. We calculate the sample error on each recorded click.
% %, and convert it into time error with a sampling rate of 15.625kHz.
% Figure ~\ref{fig:sync} shows the synchronization validation results across this experiment. We can see that with time sync enabled, the sample error never exceeds 1 sample at 15,625 kHz, or 64 $\mu$s.

% Finally we evaluate the power consumption of the ClearBuds hardware. We measure current consumption by powering our system through its Micro-USB port with a power supply, which goes through the same power path as our coin cell battery. We set our power supply voltage to 3V and set a maximum current output of 300mA, to reflect the internal 10 ohm resistor of a CR2032 battery. While streaming microphone data, we measure current consumed to be between 5-6mA. With the CR2032's nominal capacity of 210mAh, this translates to 35-42 hours of streaming  operation.

%\begin{figure}[t!]
%\vskip -0.1in
%\centering
%\includegraphics[width=1\linewidth]{figures/mic-pipeline.png}
%\caption{\small{Microphone Data Pipeline}}
%\label{fig:mic_pipeline}
%\end{figure}

%[SHOW TIME SYNC PLOT]

%Two plots: Sine wave with time sync ON and sine wave with time sync OFF
%How to best represent drift on a graph? (5, 10, 15, 30, 60min?)
\section{Training methodology} \label{sec:datasets}
%Designing a network that generalizes to in-the-wild settings requires carefully choosing the training methodology.  Training the network requires clean ground truth speech samples as supervised learning targets, which is difficult to achieve in in-the-wild settings since the groundtruth speech is corrupted with background noise and voices. Therefore, our data collection is based on the mix-and-separate framework \cite{zhao2018sound}, where clean speech and noise samples are recorded separately and combined randomly to form noisy mixtures. %The network then learns to predict the clean speech given the noisy mixture.  Here, we describe how  clean ground truth samples and noisy mixtures are generated to train the network. The network is jointly trained on all these  datasets.

Training the network in a supervised way requires clean ground truth speech samples as training targets. This is difficult to obtain in fully natural settings since the ground truth speech is corrupted with background noise and voices. Training a network that generalizes to in-the-wild scenarios also requires the training data to mimic the dynamics of real speech as closely as possible. This includes reverb, voice resonance, and microphone response. Synthetically rendered spatial data is the easiest type of data to obtain, but most different from real recordings, while real speakers wearing the headset in an anechoic chamber provide the best ground-truth training targets, but are the most costly to obtain. Synthetic data can simulate various reverb and multipath that are not captured in an anechoic chamber. We adopt a hybrid training methodology where we first train on a large amount of synthetic data and fine-tune on real data recorded with our hardware. Our training method is based on the commonly used mix-and-separate framework \cite{zhao2018sound}, where clean speech and noise samples are recorded separately and combined randomly to form noisy mixtures. Our  results  show that our network trained this way generalizes to naturally recorded noisy data in  real-world environments.

{\bf Synthetic data.}  This type of data is the easiest to obtain, since a wide variety of voice types and physical setups can be generated instantly. Many machine learning baselines, e.g., \cite{luo2020endtoend, jenrungrot2020cone, tzirakis2021multichannel}, only train and evaluate on synthetic data generated in this manner. To generate the synthetic dataset, we create multi-speaker recordings in simulated environments with reverb and background noises. All voices come from the VCTK dataset \cite{vctk} (110 unique speakers with over 44 hours), and background sounds come from the WHAM! dataset \cite{wham}, with 58 hours of recordings from a variety of noise environments such as a restaurant,  crowd,  and music.

To synthesize a single example, we create a 3 second mixture as follows: two virtual microphones are placed 17.5~cm apart, which is the average distance between human ears \cite{RISOUD2018259}. The target speaker's voice is placed  at the center between the two virtual microphones, and a second voice is placed randomly between $1$ and $5$ meters away and at a random angle. A randomly chosen background noise is also placed in the scene. We then simulate room impulse responses (RIRs) for a randomly sized room using the image source method implemented in the pyroomacoustics library \cite{allen1979image, scheibler2018pyroomacoustics}. The room is rectangular with sides randomly chosen between 5 and 20 meters, and the RT60 values are randomly chosen between 0 and 1 second. All signals are convolved with the RIR and rendered to the two channel microphone array. The volumes of the background are randomly chosen so that the input signal-to-distortion ratio is roughly between -5 and 5 dB. For training, we use 10,000 mixtures generated in this manner.

{\bf Hardware data. } While a large amount of synthetic data can be easily rendered to train the network, it does not contain characteristics such as the microphone response of physical hardware and imperfections in the time-of-arrival. To address this, we also train on a set of recorded voice samples from our earbuds. We set up a foam mannequin head with an  artificial mouth speaker (Sony SBS-XB12) that plays VCTK samples as the spoken ground truth. For background voice recordings, the speaker is placed in varying locations within a one meter radius of the foam head. Physically recorded background noise is provided by binaural version of the WHAM! dataset \cite{wham}, which was recorded in real environments using a binaural mannequin like ours. We record 2 hours each of clean speech, and background voices. 2000 random mixtures are then created for training.

{\bf Human data.} The spoken hardware data above still does not contain natural voice resonance since it is played out of an electronic speaker. Furthermore, the background sounds recorded by a mannequin wearing earbuds still misses some of the physical filtering of the human body. To better capture desired output of real scenarios, we collect a ground-truth speech dataset in an anechoic chamber with human speakers (5 male, 4 female) and a noise dataset in real environments with human listeners. For the voice data, each human speaker wore our ClearBuds prototypes, and uttered 15 minutes of text from Project Gutenberg in the anechoic chamber. The purpose of this anechoic data is to provide clean training targets for the network, modelling the resonance of human speakers wearing our hardware.
For the real world  noise dataset, individuals wore ClearBuds and recorded various noisy scenarios such as washing dishes, loud indoor/outdoor restaurants, and busy traffic intersections. 2000  random mixtures of clean voice and recorded noise were generated for this dataset. 

Our network is jointly trained using all these  datasets. Note that  testing and evaluation is  done  {\it outside} the anechoic chamber. 

% Finally, we collect an evaluation set with 2 human speakers (1 male, 1 female) in 2 noisy environments, both unseen and untrained on our network.  We also  provide  supplementary videos of the system being used in the wild in unseen locations and speakers.

%  When training, we randomly drop out the background noises and voices. In particular, each training example contains the target speaker all the time, and the background noise and background voice each with $70\%$ probability. This helps the network generalize to different background noise scenarios, including scenarios with no background noise or background voice. 

\begin{figure}
\vskip -0.1in
\centering
\includegraphics[width=1\linewidth]{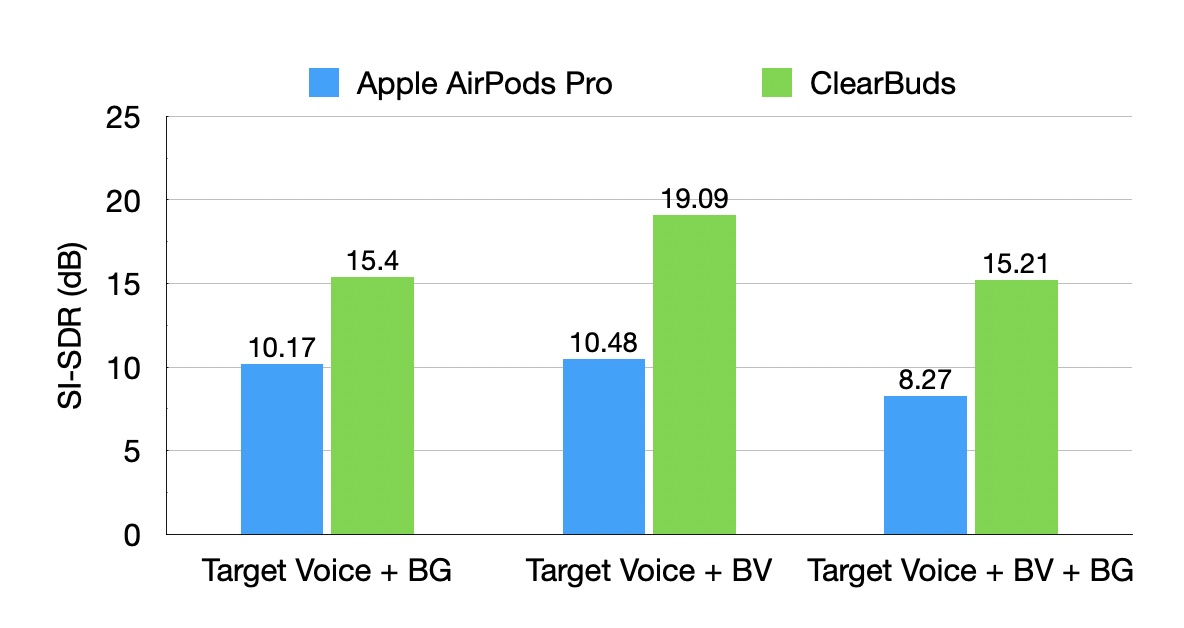}
\vskip -0.2in
\caption{Comparison with AirPods Pro. Reporting the output SI-SDR (note: not SI-SDR increase). ClearBuds exceeds in three conditions: target voice plus background noise (BG), target voice plus background voice (BV), target voice plus background voice and noise.} % Data was collected with a mannequin head with artificial mouth and array of background monitor speakers.}
\vskip -0.15in
\label{fig:airpods}
\end{figure}

\section{Experiments and Results}
%We show several evaluations to demonstrate the strength of our method. First, we conduct a user study using completely in-the-wild recordings of a speaker in a noisy environment. Then we compare against AirPods using a repeatable audio setup that is run with both ClearBuds and with AirPods. To evaluate against baseline methods, we use quantitative metrics on the test set of the synthetic data and hardware data described in section \ref{sec:datasets}. Finally, we conduct several ablation studies with the synthetic dataset to show how various factors influence our speech enhancement quality. The system latency on a variety of mobile devices is also reported.

We first compare our end-to-end system performance against a commercial  wireless earbud system. We then present in-the-wild evaluation of our system. Next, we compare numerical results against various speech enhancement baselines. Finally, we present system-level evaluations. Our work is approved by the IRB.
%To compare our model's performance against that of other networks, we perform simulated experiments on synthetically rendered data as well as data collected from our ClearBuds hardware.
% To quantitatively compare our model's performance against that of other networks, we perform simulated experiments on rendered data.
% We then , we also collect data in a repeatable acoustic environment through our ClearBuds hardware.
% Next, we collect a variety of in-the-wild data and demonstrate the advantage of our hybrid network by running a \textcolor{red}{37 participant} user study and reporting mean opinion scores.
% We conduct several studies evaluating the effect reverberation, speaker angle, and ear distance.
% \textcolor{red}{Finally, we report measure and report system factors including synchronization, latency, and power consumption.}
\subsection{Comparison with  Beamforming Earbuds} \label{sec:endtoend}

 We  evaluate our end-to-end system against the Apple AirPods Pro headset connected to a  iPhone 12 Pro in a repeatable physical set up. In our evaluation, as is typical, there is  no overlap between training and test datasets.

\begin{figure*}[t!]
\centering  
%\begin{tabular}[c]{ccc}
%\multirow{2}{*}[14pt]{
%\subfigure[]{\label{}\includegraphics[width=0.5\textwidth]{figures/MOS_study_results_v3.png}}}&
\subfigure[]{\label{}\includegraphics[width=0.24\textwidth]{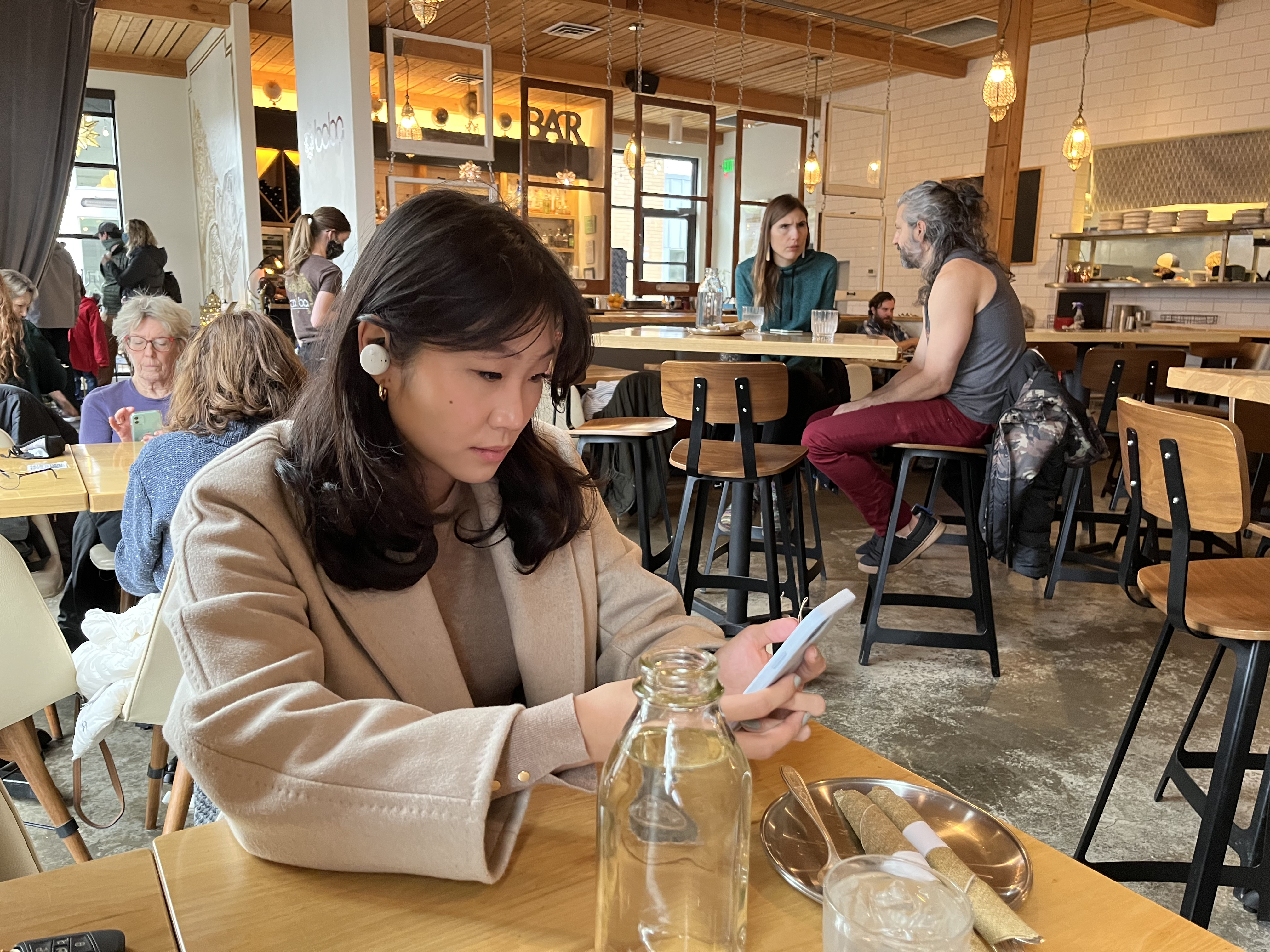}} \hfill
\subfigure[]{\label{}\includegraphics[width=0.24\textwidth]{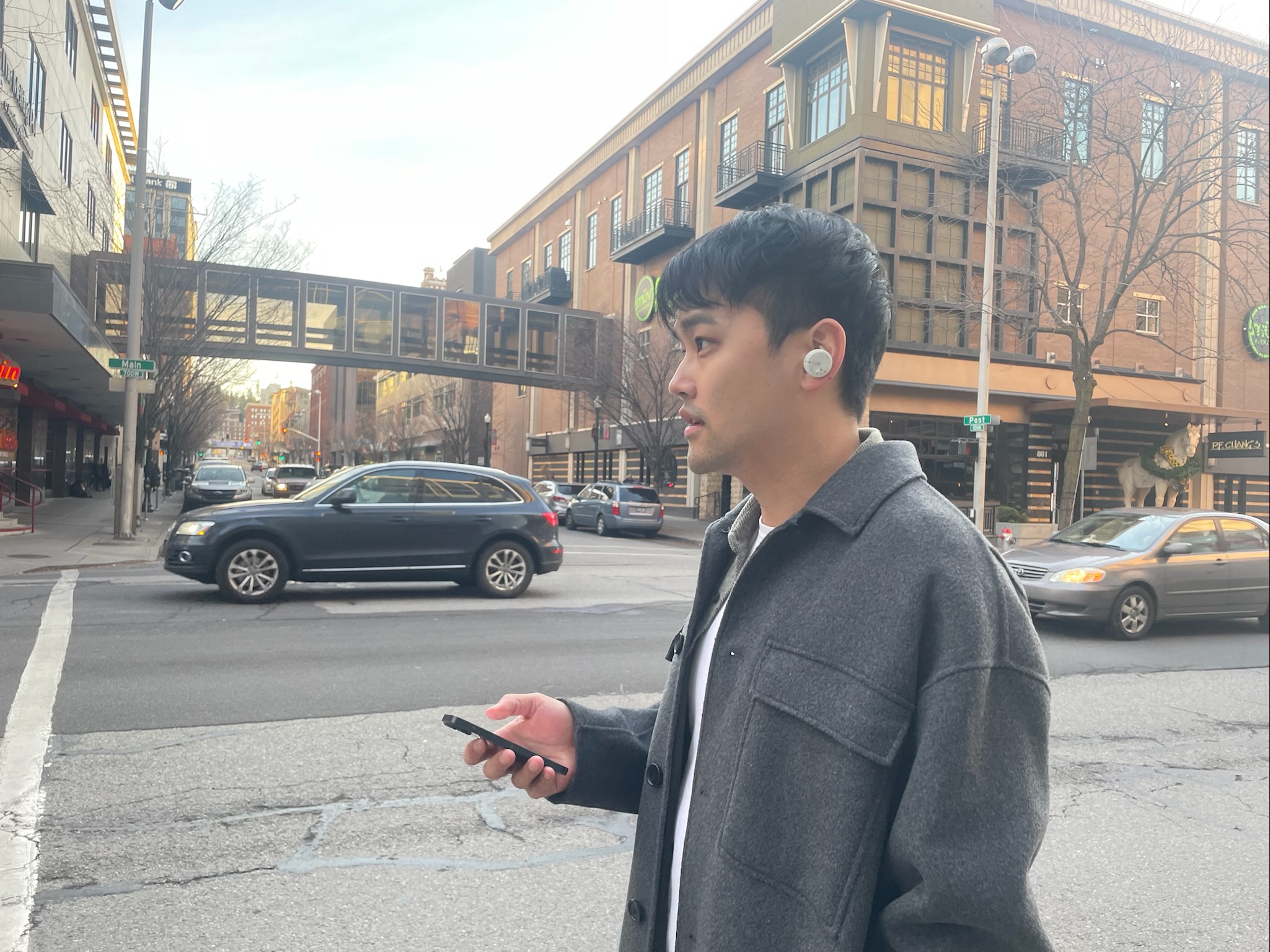}} \hfill
\subfigure[]{\label{}\includegraphics[width=0.24\textwidth]{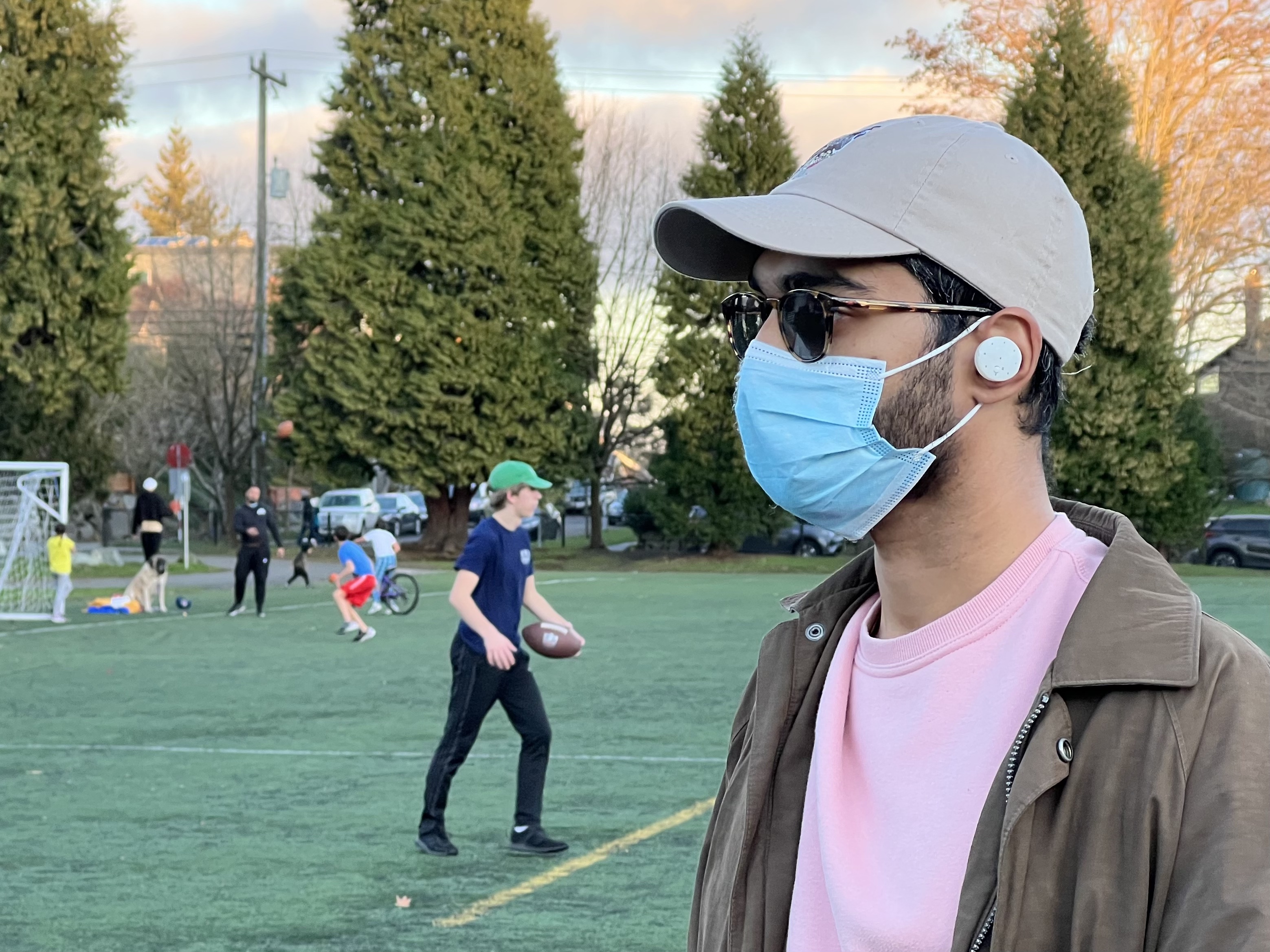}} \hfill
\subfigure[]{\label{}\includegraphics[width=0.24\textwidth]{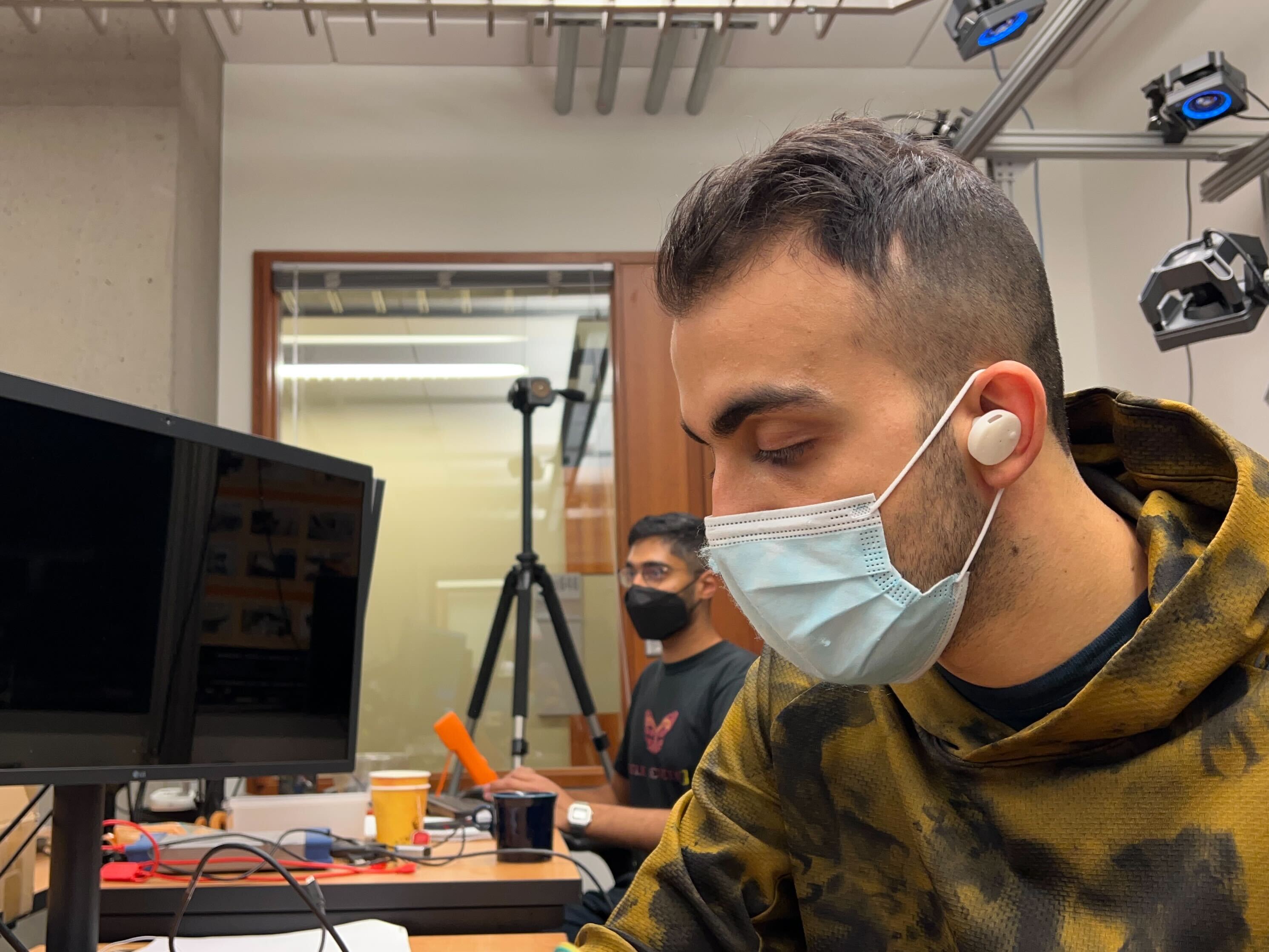}}\\
%\end{tabular}
\vskip -0.15in
\caption{{In-the-wild experiments in various  scenarios (crowded cafe, busy intersection, outdoor plaza, classroom)}  were conducted across 8 users and indoor and oudoor  environments, all unseen in our training dataset.}
\label{fig:picsenv}
\vskip -0.2in
\end{figure*}

{\bf Procedure.} We use the popular metric \textit{scale-invariant signal-to-distortion ratio} (SI-SDR)~\cite{sdr}. While SI-SDR provides a repeatable metric used in the acoustic community, it requires a clean, sample-aligned ground truth (target voice) as the basis for evaluation. Therefore, we create a repeatable soundscape for our test setup where a sample-aligned ground truth can be obtained.
A foam mannequin head with a speaker (Sony SBS-XB12) inserted into its artificial mouth uttered one hundred VCTK samples with identities and samples unseen in the training set. The mannequin wore ClearBuds and AirPods Pro in subsequent experiments, and the outputs of the two systems could be directly compared.
Ambient environmental sound (from WHAM! dataset) was played via four monitors (PreSonus Eris E3.5) positioned to fill 3 meter by 4 meter room, and background voice (also VCTK) was played from a monitor positioned 0.4 meters from head on the right. 
 %To create a repeatable soundscape for each test condition,
All speakers were driven through a common USB interface (PreSonus 1810c) ensuring the same time-alignment and loudness between the two  test conditions.
Since Apple AirPods Pro beamforming cannot be toggled on and off, we cannot calculate an SI-SDR increase (SI-SDRi), and therefore report output SI-SDR.
To establish the ground truth voice against which to calculate SI-SDR, we record clean target voice through each headset.
Ambient noise SNR ranged between 0dB and 16dB with respect to target voice. Qualitatively, this sounded like a second person speaking loudly in a noisy bar or cafe. Finally, background voice SNR ranged between 6dB and 12dB, qualitatively sounding like a person speaking from a meter or two away.

{\bf Results.}  We report output SI-SDR from the two systems in Fig.~\ref{fig:airpods}. To calculate output SI-SDR, we align individual one second chunks and take the logarithmic mean across 250 chunks. We find that ClearBuds achieves  higher output SI-SDR across all test conditions when  compared to the beamforming utilized by the Apple AirPods Pro. For a qualitative comparison of AirPods Pro versus ClearBuds performance with human speakers, see video: \textcolor{blue}{{{\url{https://clearbuds.cs.washington.edu/videos/airpods_comparison.mp4}}}}. %updated 2022-05-25

\subsection{In-the-Wild Evaluation}\label{sec:mos} 

We perform in-the-wild evaluation   in  indoor and outdoor scenarios as well as users not in the training data.
The procedure and results are described in the following sections.
% As described in section \ref{sec:nn}, in our hybrid network we apply an additional learned, low-cost spectrogram mask to clean unpleasant artifacts that surfaced when resource-constraining the Conv-TasNet architecture.
% As seen in Table \ref{table:results}, this change is not well characterized by SI-SDRi or PESQ metric as the SI-SDRi and PESQ of our CB-Conv-TasNet model did not differ from those of our hybridized, CB-Net architecture. So, we conducted a subjective listening test to comparing these conditions.

\begin{figure}[t!]
%\vskip -0.1in
\centering
\includegraphics[width=1\linewidth]{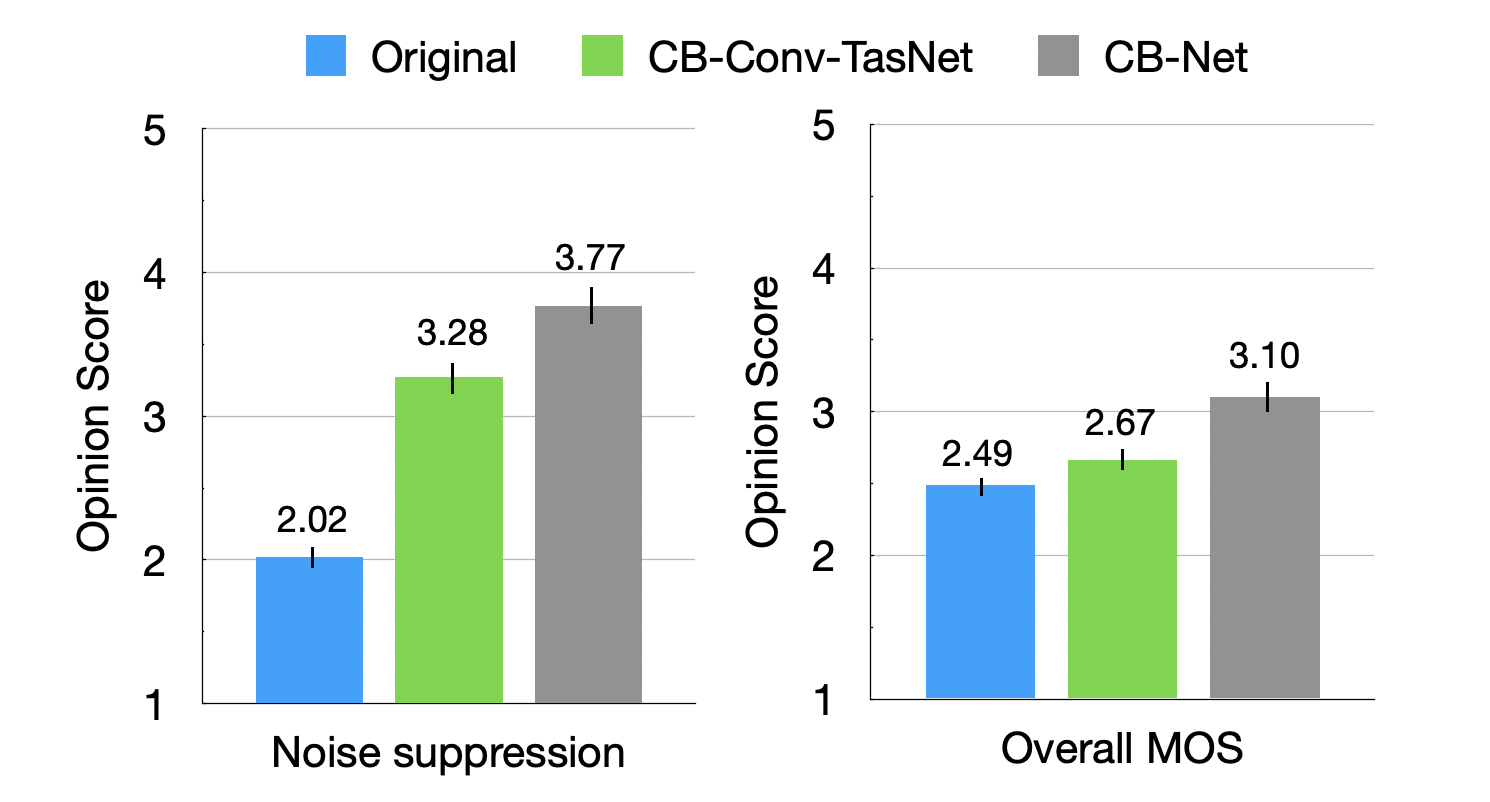}
\vskip -0.15in
\caption{In-the-wild study results. Noise suppression indicates perceived quality of background noise reduction (higher is less intrusive). Overall MOS indicates overall perceived quality. Error bars are 95\% CI.} % Data was collected with a mannequin head with artificial mouth and array of background monitor speakers.}
\vskip -0.2in
\label{fig:mos}
\end{figure}

{\bf In-the-wild experiments.} Eight individuals (four male, four female, mean age 25) with a variety of accents wore a pair of ClearBuds and read excerpts from Project Gutenberg~\cite{gutenberg}  while in four noisy environments: a coffee shop, a noisy intersection, an outdoor plaza, and a classroom (see Fig.~\ref{fig:picsenv}). The environments featured ringing phones, cross-talk from other people, ambient music, a crying baby, opening/closing doors, driving vehicles, and street noise, amongst other common sounds. These experiments  were uncontrolled in that the background voices and noise were naturally occurring sounds that are  typical to these real-world scenarios and were mobile.

{\bf Evaluation procedure.} In-the-wild evaluation precludes access to clean, sample-aligned truth to compute SI-SDR. Instead, the common (and expensive)  procedure is to perform a user study and compute the mean opinion score. Since this is a time-consuming process, prior works on binaural networks, e.g.,  \cite{luo2020endtoend, binaural_osu, jenrungrot2020cone}, avoid in-the-wild evaluation.  Since our goal is to design and evaluate an in-ear system in real scenarios, we recruit thirty-seven participants (11 female, 26 male, mean age 29)  for a user study. Each participant listened to between 6 and 11 in-the-wild audio samples (avg. 9.38 samples, each between 10--60 seconds). Each speech sample was processed and presented three ways: (1) the original input, (2) CB-Conv-TasNet, and (3) CB-Net, yielding a total of 37$\times$9.38$\times$3 $=$ 1,041 rating samples.
%Order of presentation was counter-balanced.

Participants were encouraged to use audio equipment they would typically use for a call. Fourteen used earbuds, thirteen used computer speakers, seven used headphones, and three used phone speakers. The study took about 25 minutes per participant. As is typical with noise suppression systems, participants were asked to give ratings in two categories: the intrusiveness of the noise and overall quality (mean opinion score - MOS):

\begin{small}
\begin{enumerate}
    \item \textbf{Noise suppression: }\textit{How INTRUSIVE/NOTICEABLE were the BACKGROUND sounds? 1 - Very intrusive, 2 - Somewhat intrusive, 3 - Noticeable, but not intrusive, 4 - Slightly noticeable, 5 - Not noticeable}
    \item \textbf{Overall MOS: }\textit{If this were a phone call with another person, How was your OVERALL experience? 1 - Bad, 2 - Poor, 3 - Fair, 4 - Good, 5 - Excellent}  
\end{enumerate}
\end{small}

\begin{figure}[t!]
\centering  
\subfigure[]{\label{}\includegraphics[width=0.23\textwidth]{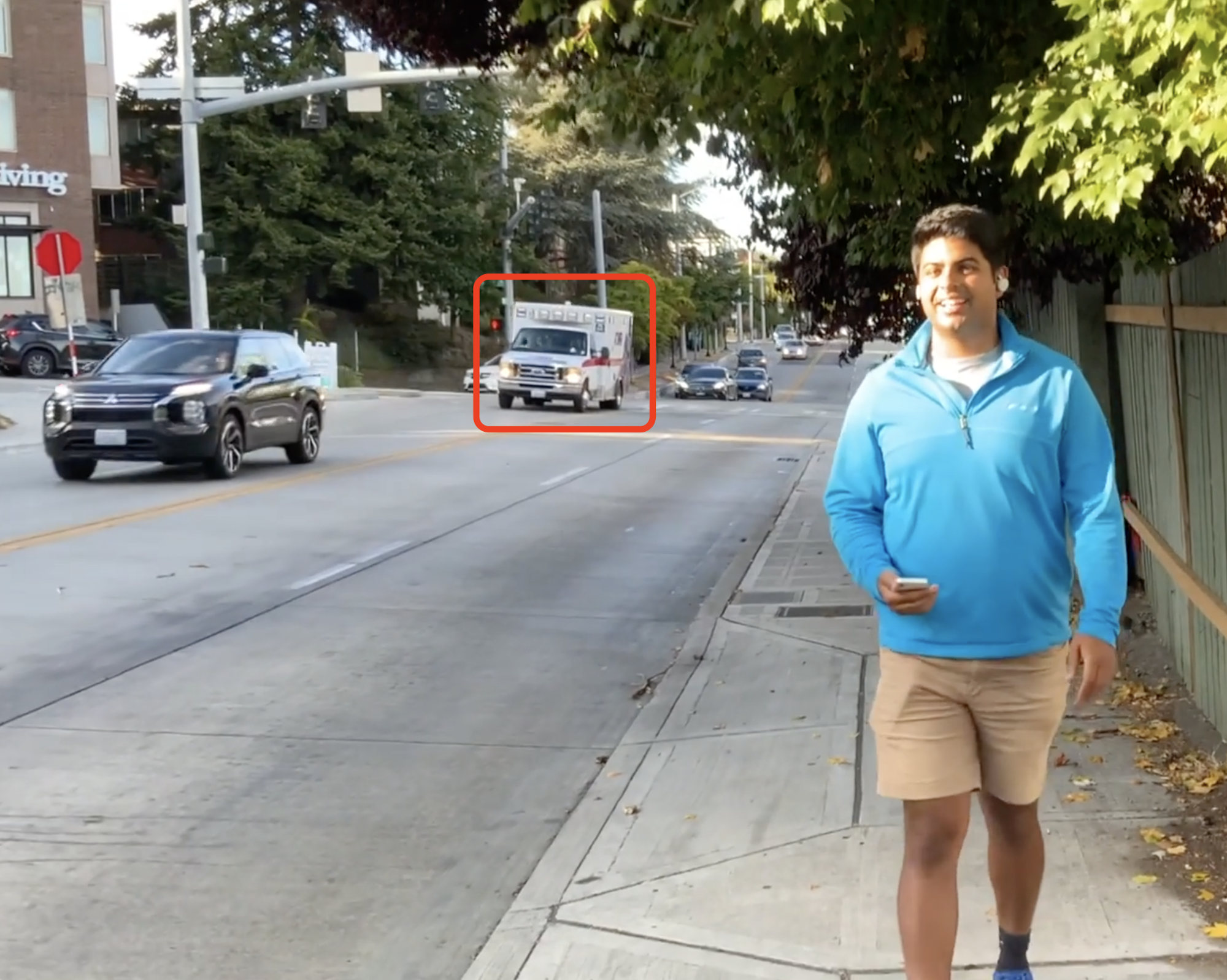}} \hfill
\subfigure[]{\label{}\includegraphics[width=0.23\textwidth]{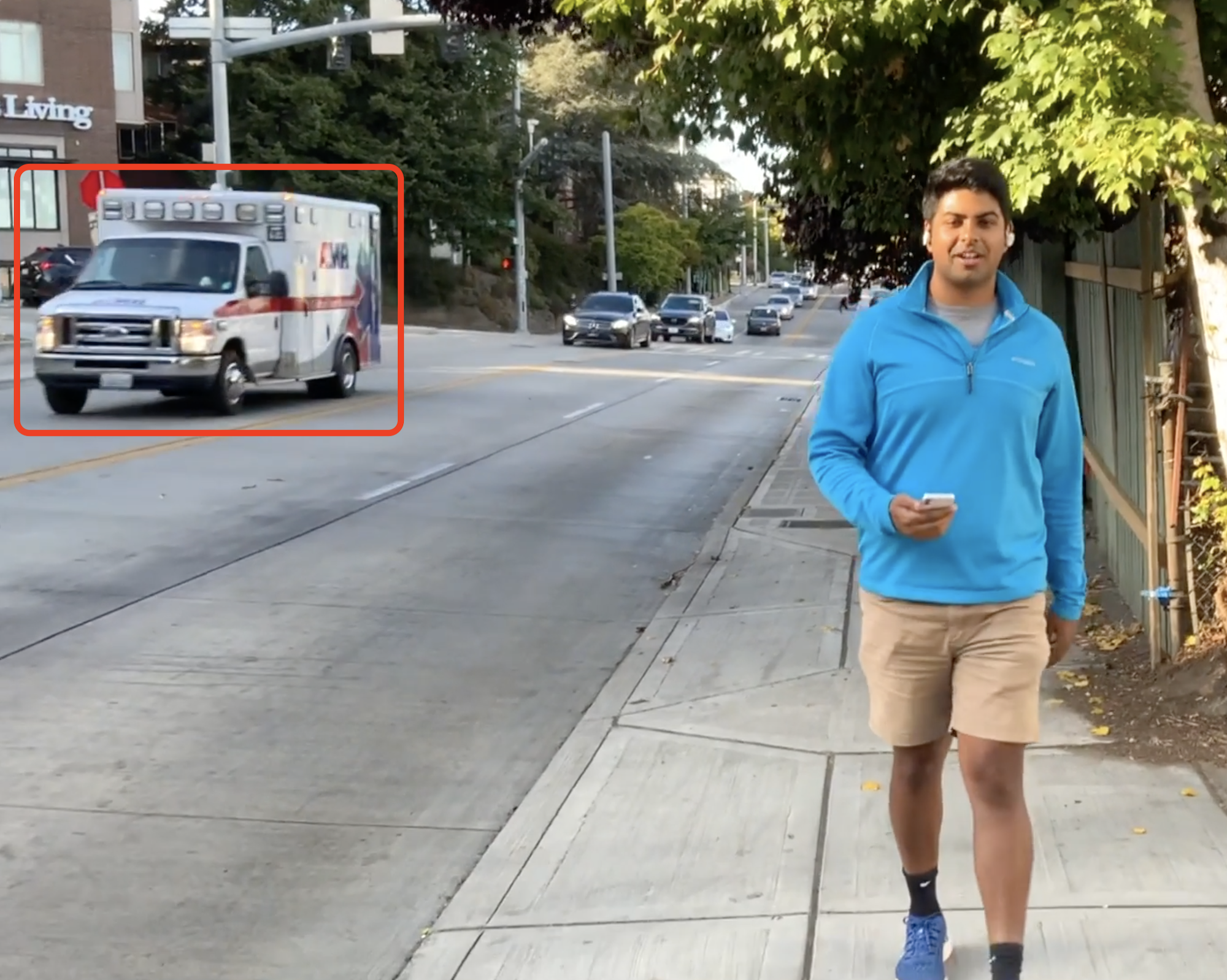}}\\
%\end{tabular}
\vskip -0.15in
\caption{Mobility of speaker and noise sources in-the-wild. Red box highlights a moving truck on the road while ClearBuds user is walking.  Video: \linebreak \textcolor{blue}{\url{https://youtu.be/HYu0ybjcQPA?t=127}}}
\label{fig:mobility}
\vskip -0.2in
\end{figure}

{\bf Results.} Fig.~\ref{fig:mos} shows the noise intrusiveness and MOS values for the original microphone, CB-Conv-TasNet, and CB-Net.
As expected, applying CB-Conv-TasNet to the original audio helped suppress noise dramatically, increasing opinion score from 2.02 (slight better than \textit{2 - Somewhat intrusive}) to 3.28 (between \textit{3 - Noticeable, but not intrusive} and \textit{4 - Slightly noticeable}) (p<0.01).
The light-touch, spectrogram-masking clean up method featured in CB-Net increased noise suppression opinion score significantly (p<0.001) to  3.77, indicating the method did indeed further suppress perceptually annoying noise artifacts.
Importantly, this step also increased overall MOS. While users only slightly preferred (p<0.05) CB-Conv-TasNet (2.67) to the original input (2.49) due to artifacts introduced, they more significantly (p<0.001) preferred our  CB-Net (3.10), an increase of 0.61 opinion score points from the input.
For context, in the  flagship ICASSP 2021 Deep Suppression Noise Challenge \cite{icaasp_dns}, with state-of-the-art, real-time algorithms run on a quad-core desktop CPU, the winning submission increased MOS by 0.57~\cite{icaasp_dns_results} from input.

\begin{table*}[t]
    \label{tab:sdr-results}
    %\vspace{-5mm}
	\centering
	\caption{ Benchmarking our neural network. We show results for a target voice speaking in three noise scenarios: (1) Background noise (BG), (2) Background voice (BV), and (3) Background noise and background voice (BG and BV).  CB-Conv-TasNet performs slightly better on  synthetic data, but as shown in Fig.~\ref{fig:mos}, does not generalize  as well to  in-the-wild  scenarios. This demonstrates the importance of evaluating  networks on real in-the-wild hardware data.}
	\small
	\begin{tabular}{c|ccc|ccc}
		\toprule
		& \multicolumn{3}{c}{SI-SDR increase (SI-SDRi)} & \multicolumn{3}{c}{Output PESQ} \\  
		\noalign{\vskip 0.3mm}    
% 		\toprule
		\midrule
		\noalign{\vskip 0.1mm}    
		Method & \shortstack{Target with\\ BG} & \shortstack{Target with\\ BV} & \shortstack{Target with\\ BV + BG} & \shortstack{Target with\\ BG} & \shortstack{Target with\\ BV} & \shortstack{Target with\\ BV + BG} \\
		\noalign{\vskip 1mm}    
		\hline
		\noalign{\vskip 1mm}    
		\textbf{CB-Net} & 10.41 & 10.56 & 9.35 & 2.08 & 2.68 & 1.81 \\
		CB-Conv-TasNet   & 11.19 & 11.01& 9.68 & 2.24 & 2.58 & 1.91 \\
		CB-Conv-TasNet Single Mic & 6.15 & 0.13 & 2.34 & 1.82 & 1.84 & 1.53 \\
 		CB-UNet & 3.21 & 0.78 & 1.82 & 1.60 & 2.10 & 1.50 \\
 		{DTLN ~\cite{DTLN}} & 7.02 & 0.06 & 2.13 & 2.08 & 1.95 & 1.67 \\
		Causal Demucs ~\cite{demucsreal} & 6.62 & -0.03 & 2.11 & 1.80 & 1.88 & 1.43 \\
		Ideal Ratio Mask (IRM, oracle) & 11.41 & 11.53 & 12.04 & 2.53 & 3.00 & 2.44 \\
		Ideal Binary Mask (IBM, oracle) & 9.97 & 11.05 & 10.85 & 2.30 & 2.90 & 2.21 \\
%		Delay-and-Sum Beamforming & 1.97 & 0.89 & 1.32 & 1.67 & 2.02 & 1.56 \\
		\bottomrule
	\end{tabular}
	\label{table:results}
\end{table*}

\begin{figure*}
\centering  
\vskip -0.15in %%% not \center
\subfigure[]{\label{fig:angle}\includegraphics[width=0.3\textwidth]{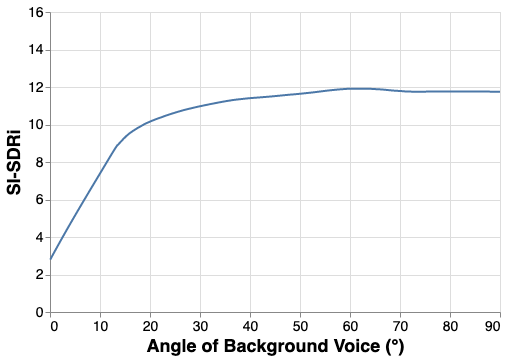}}\hfill
\subfigure[]{\label{fig:reverb}\includegraphics[width=0.3\textwidth]{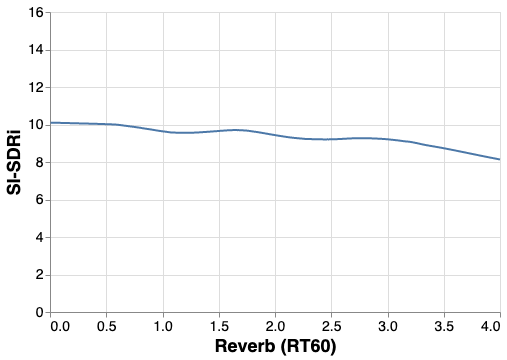}}\hfill
\subfigure[]{\label{fig:head-width}\includegraphics[width=0.3\textwidth]{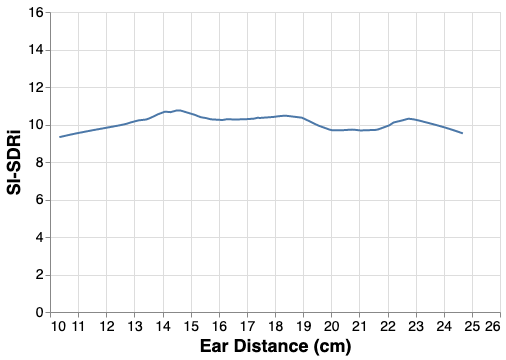}}
\vskip -0.15in
\caption{(a) Performance against angle of background voice in presence of significant multipath.  (b) Performance against amount of reverberation in an indoor room. RT60 (in seconds) measures how long sound takes to decay by 60 dB in a space with a diffuse soundfield. (c) Performance as distance between ears increases.}
\label{fig:ml-performance}
\vskip -0.2in
\end{figure*}

Note that in our in-the-wild experiments, the background noise and voices were not static. The speakers themselves can  also be  mobile (see Fig.~\ref{fig:mobility}). Our network was able to adaptively remove the background noise and achieve speech enhancement with  mobility.

\subsection{Benchmarking our Neural Network}
The conventional evaluation in the machine learning and acoustic community is to evaluate models and techniques on synthetic data against baselines. For completeness, we compare our method against a variety of speech enhancement baselines using the synthetic dataset. For evaluation, an additional 1000 mixtures of 3 seconds each were generated such that there was no overlapping identities or samples between the train and test splits.

{\bf Evaluation Procedure.} For  comparisons to other baseline methods, we use the popular 
%the popular metrics \textit{scale-invariant signal-to-distortion ratio} (
SI-SDR  and {PESQ} metrics.
Unlike the AirPods experiment, where the original noisy mixture could not be recorded since AirPods beamforming cannot be toggled off, here we compute SI-SDR of the ground truth relative to both the input noisy mixture and then to the network output. When reporting the increase from the input SI-SDR to output SI-SDR, we use the  SI-SDR improvement (SI-SDRi).
%Unlike the AirPods experiment, where the noisy mixture could not be recorded, the SI-SDR of the ground truth here is computed relative to the noisy mixture and then to the network output. When reporting the increase from the input to output SI-SDR, we use the  SI-SDR improvement (SI-SDRi).

For a deep learning baseline in the waveform domain, we choose the causal Demucs model~\cite{demucsreal}. This is a single channel method which was recently shown to outperform many other deep learning baselines and runs real-time on a laptop CPU. We also compare with Dual-signal Transformation LSTM Network (DTLN) ~\cite{DTLN}. This method also runs on a laptop or mobile phone in real-time. To compare with spectrogram based methods, we use the oracle baselines, ideal ratio mask (IRM) and ideal binary mask (IBM) \cite{sigsep, wang2005ideal}, that use the ground truth voice to calculate the best possible result that can be obtained by masking a noisy spectrogram. 

As an ablation study, we report results with each individual component of the network,  \textit{CB-Conv-TasNet} and \textit{CB-UNet}. We also show results when the multi-channel part of our network, \textit{CB-Conv-TasNet}, only has access to one microphone, labeled as \textit{CB-Conv-TasNet Single Mic}. This explicitly shows the advantage of using two microphones.  There are only a few deep learning methods that tackle binaural speech separation for mobile processing, and the most relevant ones, such as \cite{binaural_osu} and \cite{han2020realtime}, do not have publicly available code to test against.

{\bf Results.} As shown in Table \ref{table:results}, our binaural method is comparable to the best possible results that can be obtained by a spectrogram masking method (IBM, IRM). We also show an improvement over waveform based deep learning methods that only use a single microphone input. In particular, the improvement is greatest when there are two speakers present (Target Voice + Background Voice). This is because single channel methods can only rely on voice characteristics, whereas our network also uses spatial cues to separate the speaker of interest. Although \textit{CB-Net} shows similar or worse performance to \textit{CB-Conv-TasNet}, subjective evaluation on in-the-wild hardware data shows that \textit{CB-Net} is far superior to human listeners (see~\xref{sec:mos}).

Examples of the synthetic dataset, outputs from all the methods and  qualitative comparisons against Krisp \cite{krisp}, a commercial noise suppression system, can be found linked from our project website: 
\textcolor{blue}{{{\url{https://clearbuds.cs.washington.edu}}}}.
%\textcolor{blue}{{{\url{https://bit.ly/3yJ6DwJ}}}}

%Further, to demonstrate our hybridized network's subjective background reduction and speech quality aspects, we run a  user study to compare the noise and speech performance amongst our mobile-deployed network architectures.

\subsubsection{Additional neural network  evaluations} We numerically evaluate various aspects of the design by changing the angle of background voice,  reverberance in the  environment, and microphone separation.

{\bf Angle of background voice.} The ability of our network to separate the target voice from a background voice is based on utilizing the time difference of arrival to the binaural microphones. Because we only have two microphones, this ability is limited when the background voice is in the front-back plane of the speaker. In this case, the background voice will arrive at each microphone simultaneously, and there will be no spatial cues to separate the two voices.  To illustrate this effect, we graph the separation performance as a function of the angle of the background voice in Fig.~\ref{fig:angle}.

{\bf Multipath and reverberant environments.} While our in-the-wild experiments show the performance  in various indoor and outdoor environments, we benchmark our system  in different  reverberant conditions, including those more reverberant than seen during training. Synthetically generated mixtures are generated using the pyroomacoustics library  with the RT60 value randomly chosen between 0 and 4s. We generate 200 examples and plot the SI-SDRi compared to the RT60 in Fig.~\ref{fig:reverb}. Our method shows only a slight decrease in performance as the reverberation of the environment increases. Because the target speaker is physically close to the microphone array, our setup is generally less affected by reverberations than other kinds of source separation problems where the target speaker may be further away.

\begin{figure}[t!]
\centering     %%% not \center
\subfigure{\label{fig:sync-a}\includegraphics[width=0.36\textwidth]{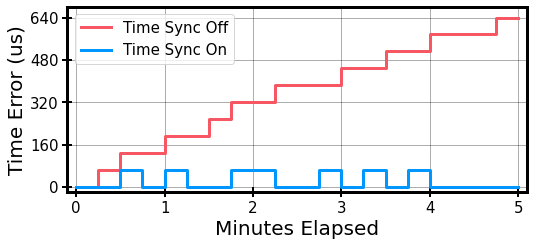}}\vskip -0.12in
\subfigure{\label{fig:sync-b}\includegraphics[width=0.36\textwidth]{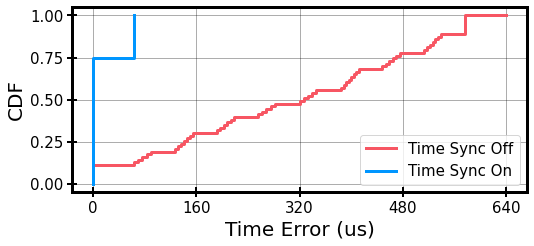}}
\vskip -0.15in
\caption{Time Synchronization Validation.  Without time synchronization (red), microphone samples drift apart and lose alignment at about 128$\mu$s/min.} %Also shows  CDF across three experiments of 5 minutes each. }
\label{fig:sync}
\vskip -0.2in
\end{figure}

{\bf Separation between microphones.} Our in-the-wild evaluation across 8 participants showed  generalization across facial features. Here, we  benchmark our method  to different head sizes where the distance between the microphones may be different. We generate 200 synthetic samples, where the distance between the microphones is randomly chosen between 10 and 25~cm. Because the target speaker is in the middle of the microphone array, the target signal will arrive at both mics simultaneously regardless of the microphone distance.  Fig.~\ref{fig:head-width} show little change in performance even with microphone distances greatly different than used during training.

\subsection{System   Evaluation}
%We evaluate various system parameters like synchronization, latency and power consumption. 
{\bf Synchronization.}  In order to evaluate this, we place both ClearBuds roughly equidistant from a speaker. A {click tone} is played every 15 seconds for 5 minutes, and recorded on both ClearBuds with time sync disabled and enabled. We calculate the sample error on each recorded click offline and convert it into time error with a sampling rate of 15.625kHz.
Fig.~\ref{fig:sync}(a) shows the synchronization  results across a five minute interval. With time sync enabled, the sample error never exceeds 1 sample at 15,625 kHz, or 64 $\mu$s. Fig.~\ref{fig:sync}(b) also shows the CDF of the timing error across  experiments of 5 minutes each conducted with other Bluetooth devices in the environment,  with and without time synchronization.

{\bf Run-time and end-to-end latency.} Mouth-to-ear delay is defined as the time it takes from speech to exit the speaker's mouth and reach the listener's ear on the other end of the call. The International Telecommunication Union Telecommunication Standardization Sector (ITU-T) G.114 recommendation regarding mouth-to-ear delay indicates that most users are ``very satisfied'' as long as the latency does not exceed 200~ms \cite{g.114}. In our end-to-end system, we targeted a one-way latency of 100~ms prior to uplink, leaving up to 100~ms of network delay to move an IP packet from the source to the destination.

With a 180-sample PCM buffer being filled at 31.25 kHz, there is a 5.76~ms delay prior to the samples reaching BLE stack. Once these samples reach the radio hardware, there is a worst-case additional latency of 7.5~ms as defined by the minimum BLE connection interval supported by Bluetooth 5.0 \cite{bt5}. At the time of writing, the latest iOS supports a minimum BLE connection interval of 15~ms. After the samples reach the mobile phone, we wait for 67.2~ms to receive enough samples to run a forward pass of our network. Our network has a run-time of 21.4~ms on an iPhone 12 Pro (see Table~\ref{tab:latency}). The number of FLOPs is  computed over each packet of 350 samples. Together, we have a  latency of 109~ms, leaving 91 ms for  one-way network delay  (RTT=182ms).

{\bf Power analysis.} {CB-Net uses   an order of magnitude lower FLOPs per second compared to Conv-TasNet on the smartphone, significantly reducing the computational and corresponding power consumption. } We also  measure  the power consumption of the ClearBuds hardware. We measure current consumption by powering our system through its Micro-USB port with a DC power supply set to 3V, which goes through the same power path as our coin cell battery.
%We set our power supply voltage to 3V and set a maximum current output of 300mA, to reflect the internal 10 ohm resistance of a CR2032 battery.
While continuously wirelessly streaming microphone data, we measure average current consumed to be 5~mA. With the CR2032's nominal capacity of 210~mAh, this translates to approximately 42 hours of operation. Table \ref{tab:hardware-power-consumption} shows a breakdown by component of the system's power consumption. The accelerometer (BMA400) and flash (W25N01GVZEIG) are omitted as they are power gated during  streaming. 

%; mobile GPS usage reduces the battery-life by couple of hours

%\textcolor{red}{We also  ran our network on an  iPhone 12 Pro continuously after disabling all other applications except for Bluetooth and found that it took XXX hours to drain the battery on the phone starting at full charge.}

\begin{table}[t]
	\centering
	\small
	\caption{ Neural network run time on smartphones }
	\begin{tabular}{ccccc}
		\toprule
		Device &  Conv-TasNet & CB-Conv-TasNet & \textbf{CB-Net} \\
		\noalign{\vskip 0.5mm} 
		\hline
		\noalign{\vskip 1mm} 
		iPhone 12 Pro & 155.5ms & 17.5ms & 21.4ms \\
		iPhone 11 & 165.4ms & 18.6ms & 22.7ms \\
		iPhone XS & 241.5ms & 27.2ms &  33.0ms \\
		\hline
		\noalign{\vskip 1mm} 
		FLOPs/packet & 1078M & 97M & 131M \\
% 		iPhone 12 Pro & 155.5ms & 17.5ms & 3.9ms & 21.4ms \\
% 		iPhone 11 & 165.4ms & 18.6ms & 4.1ms & 22.7ms \\
% 		iPhone XS & 241.5 & 27.2ms & 5.8ms &  33.0ms \\
		\bottomrule
	\end{tabular}
	\vskip -0.1in
	\label{tab:latency}
	%\vspace{-8mm}
	%\vskip -0.1in
\end{table}

% \section{Discussion}
% \begin{todo-env}
% Because Krisp is a mono-input method, we expect that our binaural method would work better by leveraging the additional geometric constraint. The numerical results show that the extra microphone does indeed result in improved performance. In addition, Krisp is likely trained on far more data, which explains the close performance. Overall Krisp works extremely well for background noise removal, but there is some room for improvement. For example, Krisp can only perform audio denoising not true audio separation of multiple voices. If there are other speakers in the scene, Krisp will let them both through or sometimes play the wrong one. Because our method is spatially grounded, it can separate a single speaker from other speakers, even when the target is not the loudest. This result shows the promise of using multi-microphone methods over commercially available single channel methods.

% \citet{defossez2019music}
% Finally, looking toward wearable deployment, we look forward to training and implementing the model on data collected with our custom hardware system, Shio, in real-world scenarios.

% \end{todo-env}
\section{Limitations \& Future work}
% While ClearBuds presents state-of-the-art speaker separation in noisy environments, there are a few limitations to note for future work. 
The first limitation is that the user must be wearing both wireless earbuds to benefit from our binaural noise suppression network. Second, with only two microphones, there is an opportunity for background voices to remain in the uplink channel if the voice is within a few degrees of the target speaker's sagittal plane (see Fig.~\ref{fig:angle}). The underlying assumption of our network is that the mouth is in the middle of the user's ears, though as seen in Fig. \ref{fig:head-width} and our in-the-wild evaluation, some variance is permissible.

While we minimize the power consumption of the ClearBud hardware, we shift the processing and therefore power consumption to the more powerful mobile phone.  Performing network computation on the mobile phone over a cloud GPU is an enhancement in terms of user privacy and security so that sensitive voice data is not transmitted to the cloud.   While mobile chips are becoming more power efficient, an alternative design to explore is to run our neural network  on a plugged-in edge device (e.g., router),  minimizing computation while achieving similar latency.

{Future work could integrate two microphones in each earbud, so that each earbud could beamform toward the user's mouth prior to processing in the neural network. We also had to develop a custom wireless audio protocol to stream two microphones to a single phone. While this prevents this architecture from being deployed on today's commodity wireless earbuds, adoption may be imminent as Bluetooth 5.2 shows promise with the introduction of Multi-Stream Audio and Audio Broadcast \cite{le-audio-faqs}.}

{ Our network could also be deployed on other multi-microphone mobile or resource-constrained edge systems such as smart watches, augmented reality glasses, or smart speakers to allow for enhanced voice control or telephony in noisy environments. The hardware and firmware for Clearbuds could be leverage to produce wireless, synchronized microphone arrays for telephony, acoustic activity recognition or for swarm robot localization and control.} 

%But,  running the network on the GPU of the mobile phone comes at the cost of its  battery life.

\section{Conclusion}
Real-time speech enhancement has been an open research challenge for multiple decades. The recent proliferation of wireless earbuds and  neural network architectures provides an opportunity to build  systems that bridge neural networks and wireless earbuds to create new capabilities. Here, we present ClearBuds, the first deep learning based system to achieve real-time speech enhancement with binaural wireless earbuds. At its core is a new open-source wireless earbud design capable of operating as a synchronized binaural microphone array and a lightweight cascaded neural network. In-the-wild  experiments show that ClearBuds can achieve background noise suppression, background speech removal, and speaker separation using wireless earbuds. 

% Cite: https://www.twilio.com/blog/understanding-latency
% Cite: https://www.silabs.com/community/wireless/bluetooth/knowledge-base.entry.html/2015/08/06/audio_latency_withb-7BGG

% https://towardsdatascience.com/real-time-noise-suppression-using-deep-learning-38719819e051

\vskip 0.05in\noindent{\bf Acknowledgments.}
This research is funded by  the UW Reality Lab, Moore Inventor Fellow award \#10617 and the researchers are also funded by the National Science Foundation. We thank our shepherd, Youngki Lee, and the anonymous reviewers for their feedback on our submission.

\begin{table}[t]
	\centering
	\caption{ ClearBuds hardware power consumption }
	\small
	\begin{tabular}{cc}
		\toprule
		\bf{Component} & \bf{Power Consumption} \\
		\noalign{\vskip 0.5mm} 
		\hline
		\noalign{\vskip 1mm} 
		BLE SoC (nRF52840) & 12.02 \it{mW} \\
		Microphone (ICS-41350) & 0.77 \it{mW} \\
		Ideal Diode (LM66100DCKT) & 0.27 \it{$\mu$W} \\
		Buck Efficiency Loss (MAX38640) & 1.75 \it{mW} \\
		\hline
		\noalign{\vskip 0.5mm} 
		\bf{Total} & 14.54 \it{mW} \\

		\bottomrule
	\end{tabular}
	\vskip -0.15in
	\label{tab:hardware-power-consumption}
%	\vskip -0.15in
\end{table}

\balance
\bibliographystyle{unsrt}
\bibliography{references}

\begin{thebibliography}{10}

\bibitem{airpodssales}
https://appleinsider.com/articles/21/03/30/apple-airpods-beats-dominated-audio-wearable-market-in-2020.

\bibitem{van1988beamforming}
Barry~D Van~Veen and Kevin~M Buckley.
\newblock Beamforming: A versatile approach to spatial filtering.
\newblock {\em IEEE assp magazine}, 5(2):4--24, 1988.

\bibitem{krim1996two}
Hamid Krim and Mats Viberg.
\newblock Two decades of array signal processing research: the parametric
  approach.
\newblock {\em IEEE signal processing magazine}, 13(4):67--94, 1996.

\bibitem{chhetri2018multichannel}
Amit Chhetri, Philip Hilmes, Trausti Kristjansson, Wai Chu, Mohamed Mansour,
  Xiaoxue Li, and Xianxian Zhang.
\newblock Multichannel audio front-end for far-field automatic speech
  recognition.
\newblock In {\em 2018 EUSIPCO}, pages 1527--1531. IEEE, 2018.

\bibitem{subakan2021attention}
Cem Subakan, Mirco Ravanelli, Samuele Cornell, Mirko Bronzi, and Jianyuan
  Zhong.
\newblock Attention is all you need in speech separation, 2021.

\bibitem{luo2019conv}
Yi~Luo and Nima Mesgarani.
\newblock Conv-tasnet: Surpassing ideal time--frequency magnitude masking for
  speech separation.
\newblock {\em IEEE/ACM Transactions on Audio, Speech, and Language
  Processing}, 2019.

\bibitem{krisp}
www.krisp.ai.

\bibitem{airpods}
Apple airpods. https://www.apple.com/airpods/.

\bibitem{esense-1}
Fahim Kawsar, Chulhong Min, Akhil Mathur, and Alessandro Montanari.
\newblock Earables for personal-scale behavior analytics.
\newblock {\em IEEE Pervasive Computing}, 17(3):83--89, 2018.

\bibitem{g.114}
{\em Series G: Transmission Systems and Media, Digital Systems and Networks},
  2003.

\bibitem{samsungglobalnewsroom_2014}
Galaxy s5 explained: Audio.
  https://news.samsung.com/global/galaxy-s5-explained-audio, Jun 2014.

\bibitem{sennheiser_2020}
Earbuds that put sound first.
  https://en-de.sennheiser.com/newsroom/earbuds-that-put-sound-first, Mar 2020.

\bibitem{beamforming-app-note}
Microphone array beamforming.
  https://invensense.tdk.com/wp-content/uploads/2015/02/microphone-array-beamforming.pdf.

\bibitem{amazon}
Echo (3rd gen). https://www.amazon.com/all-new-echo/dp/b07nftvp7p.

\bibitem{InvenSense}
InvenSense.
\newblock Microphone array beamforming.
\newblock Technical Report AN-1140-00, InvenSense Inc., 1745 Technology Drive,
  San Jose, CA 95110 U.S.A, December 2013.

\bibitem{bluetooth}
Bluetooth~Audio Telephony and Automotive~Working Group.
\newblock Hands-free profile: Bluetooth® profile specification.
\newblock Technical Report v1.8, Bluetooth SIG, Apr 2020.

\bibitem{frost1972MVDR}
Otis~Lamont Frost.
\newblock An algorithm for linearly constrained adaptive array processing.
\newblock {\em Proceedings of the IEEE}, 60(8):926--935, 1972.

\bibitem{zhang2017deep}
Xueliang Zhang and DeLiang Wang.
\newblock Deep learning based binaural speech separation in reverberant
  environments.
\newblock {\em IEEE/ACM transactions on audio, speech, and language
  processing}, 25(5), 2017.

\bibitem{brandstein2001microphone}
Michael Brandstein.
\newblock {\em Microphone arrays: signal processing techniques and
  applications}.
\newblock Springer Science \& Business Media, 2001.

\bibitem{realtimenoise}
Yong Xu, Jun Du, Li-Rong Dai, and Chin-Hui Lee.
\newblock A regression approach to speech enhancement based on deep neural
  networks.
\newblock {\em IEEE/ACM Transactions on Audio, Speech, and Language
  Processing}, 23(1):7--19, 2015.

\bibitem{Mohammadiha_2013}
Nasser Mohammadiha, Paris Smaragdis, and Arne Leijon.
\newblock Supervised and unsupervised speech enhancement using nonnegative
  matrix factorization.
\newblock {\em IEEE Transactions on Audio, Speech, and Language Processing},
  21(10):2140–2151, 2013.

\bibitem{online_nonnegative}
Zhiyao Duan, {Gautham J.} Mysore, and Paris Smaragdis.
\newblock Speech enhancement by online non-negative spectrogram decomposition
  in non-stationary noise environments.
\newblock INTERSPEECH 2012, pages 594--597.

\bibitem{nikzad2020deep}
Mohammad Nikzad, Aaron Nicolson, Yongsheng Gao, Jun Zhou, Kuldip~K. Paliwal,
  and Fanhua Shang.
\newblock Deep residual-dense lattice network for speech enhancement.
\newblock 2020.

\bibitem{choi2019phaseaware}
Hyeong-Seok Choi, Jang-Hyun Kim, Jaesung Huh, Adrian Kim, Jung-Woo Ha, and
  Kyogu Lee.
\newblock Phase-aware speech enhancement with deep complex u-net.
\newblock 2019.

\bibitem{lstm_speechenhancement}
Felix Weninger, Hakan Erdogan, Shinji Watanabe, Emmanuel Vincent, Jonathan
  Roux, John~R. Hershey, and Bj\"{o}rn Schuller.
\newblock Speech enhancement with lstm recurrent neural networks and its
  application to noise-robust asr.
\newblock LVA/ICA 2015, page 91–99. Springer-Verlag, 2015.

\bibitem{fu2019metricgan}
Szu-Wei Fu, Chien-Feng Liao, Yu~Tsao, and Shou-De Lin.
\newblock Metricgan: Generative adversarial networks based black-box metric
  scores optimization for speech enhancement.
\newblock 2019.

\bibitem{TFMasking}
Meet~H. Soni, Neil Shah, and Hemant~A. Patil.
\newblock Time-frequency masking-based speech enhancement using generative
  adversarial network.
\newblock In {\em ICASSP 2018}.

\bibitem{germain2018speech}
Francois~G. Germain, Qifeng Chen, and Vladlen Koltun.
\newblock Speech denoising with deep feature losses.
\newblock 2018.

\bibitem{pascual2017segan}
Santiago Pascual, Antonio Bonafonte, and Joan Serrà.
\newblock Segan: Speech enhancement generative adversarial network.
\newblock 2017.

\bibitem{demucsreal}
Alexandre Defossez, Gabriel Synnaeve, and Yossi Adi.
\newblock Real time speech enhancement in the waveform domain.
\newblock 2020.

\bibitem{macartney2018improved}
Craig Macartney and Tillman Weyde.
\newblock Improved speech enhancement with the wave-u-net.
\newblock 2018.

\bibitem{googlemeet}
meet.google.com.

\bibitem{tinylstm}
Igor Fedorov, Marko Stamenovic, Carl Jensen, Li-Chia Yang, Ari Mandell, Yiming
  Gan, Matthew Mattina, and Paul~N. Whatmough.
\newblock Tinylstms: Efficient neural speech enhancement for hearing aids.
\newblock {\em Interspeech 2020}, Oct 2020.

\bibitem{yoshioka2018multi}
Takuya Yoshioka, Hakan Erdogan, Zhuo Chen, and Fil Alleva.
\newblock Multi-microphone neural speech separation for far-field multi-talker
  speech recognition.
\newblock In {\em 2018 IEEE International Conference on Acoustics, Speech and
  Signal Processing (ICASSP)}, pages 5739--5743. IEEE, 2018.

\bibitem{chen2018multi}
Zhuo Chen, Xiong Xiao, Takuya Yoshioka, Hakan Erdogan, Jinyu Li, and Yifan
  Gong.
\newblock Multi-channel overlapped speech recognition with location guided
  speech extraction network.
\newblock In {\em 2018 IEEE Spoken Language Technology Workshop (SLT)}, pages
  558--565. IEEE, 2018.

\bibitem{gu2020enhancing}
Rongzhi Gu, Shi-Xiong Zhang, Lianwu Chen, Yong Xu, Meng Yu, Dan Su, Yuexian
  Zou, and Dong Yu.
\newblock Enhancing end-to-end multi-channel speech separation via spatial
  feature learning.
\newblock {\em arXiv preprint arXiv:2003.03927}, 2020.

\bibitem{tzirakis2021multichannel}
Panagiotis Tzirakis, Anurag Kumar, and Jacob Donley.
\newblock Multi-channel speech enhancement using graph neural networks.
\newblock 2021.

\bibitem{jenrungrot2020cone}
Teerapat Jenrungrot, Vivek Jayaram, Steve Seitz, and Ira
  Kemelmacher-Shlizerman.
\newblock The cone of silence: Speech separation by localization.
\newblock 2020.

\bibitem{binaural1}
Xingwei Sun, Risheng Xia, Junfeng Li, and Yonghong Yan.
\newblock A deep learning based binaural speech enhancement approach with
  spatial cues preservation.
\newblock In {\em ICASSP 2019}, pages 5766--5770.

\bibitem{han2020realtime}
Cong Han, Yi~Luo, and Nima Mesgarani.
\newblock Real-time binaural speech separation with preserved spatial cues.
\newblock 2020.

\bibitem{li2011two}
Junfeng Li, Shuichi Sakamoto, Satoshi Hongo, Masato Akagi, and Y{\^o}iti
  Suzuki.
\newblock Two-stage binaural speech enhancement with wiener filter for
  high-quality speech communication.
\newblock {\em Speech Communication}, 53(5):677--689, 2011.

\bibitem{reindl2010speech}
Klaus Reindl, Yuanhang Zheng, and Walter Kellermann.
\newblock Speech enhancement for binaural hearing aids based on blind source
  separation.
\newblock In {\em 2010 4th International Symposium on Communications, Control
  and Signal Processing (ISCCSP)}, pages 1--6. IEEE, 2010.

\bibitem{van2008binaural}
Richard van Hoesel, Melanie B{\"o}hm, J{\"o}rg Pesch, Andrew Vandali, Rolf~D
  Battmer, and Thomas Lenarz.
\newblock Binaural speech unmasking and localization in noise with bilateral
  cochlear implants using envelope and fine-timing based strategies.
\newblock {\em The Journal of the Acoustical Society of America},
  123(4):2249--2263, 2008.

\bibitem{lyon1983computational}
Richard Lyon.
\newblock A computational model of binaural localization and separation.
\newblock In {\em ICASSP'83. IEEE International Conference on Acoustics,
  Speech, and Signal Processing}, volume~8, pages 1148--1151. IEEE, 1983.

\bibitem{kock1950binaural}
WE~Kock.
\newblock Binaural localization and masking.
\newblock {\em The Journal of the Acoustical Society of America},
  22(6):801--804, 1950.

\bibitem{binaural_osu}
Ke~Tan, Xueliang Zhang, and DeLiang Wang.
\newblock Real-time speech enhancement using an efficient convolutional
  recurrent network for dual-microphone mobile phones in close-talk scenarios.
\newblock In {\em ICASSP 2019}, pages 5751--5755, 2019.

\bibitem{dual_phone}
Ke~Tan, Xueliang Zhang, and Deliang Wang.
\newblock Deep learning based real-time speech enhancement for dual-microphone
  mobile phones.
\newblock {\em IEEE/ACM Transactions on Audio, Speech, and Language
  Processing}, pages 1--1, 2021.

\bibitem{binauralphone}
Nikhil Shankar, Gautam Shreedhar~Bhat, and Issa Panahi.
\newblock Efficient two-microphone speech enhancement using basic recurrent
  neural network cell for hearing and hearing aids.
\newblock {\em The Journal of the Acoustical Society of America}, 148:389--400,
  07 2020.

\bibitem{hybridbeam}
Anran Wang, Maruchi Kim, Hao Zhang, and Shyamnath Gollakota.
\newblock Hybrid neural networks for on-device directional hearing, AAAI 2022.

\bibitem{oesense21}
Dong Ma, Andrea Ferlini, and Cecilia Mascolo.
\newblock {\em OESense: Employing Occlusion Effect for in-Ear Human Sensing},
  page 175–187.
\newblock 2021.

\bibitem{esense-2}
Chulhong Min, Akhil Mathur, and Fahim Kawsar.
\newblock Exploring audio and kinetic sensing on earable devices.
\newblock WearSys '18, page 5–10, 2018.

\bibitem{plat-1}
Jovan Powar and Alastair~R. Beresford.
\newblock A data sharing platform for earables research.
\newblock In {\em Proceedings of the 1st International Workshop on Earable
  Computing}, EarComp'19, page 30–35, New York, NY, USA, 2019.

\bibitem{romit-1}
Zhijian Yang and Romit~Roy Choudhury.
\newblock Personalizing head related transfer functions for earables.
\newblock SIGCOMM '21, page 137–150, New York, NY, USA, 2021.

\bibitem{infection}
Justin Chan, Sharat Raju, Rajalakshmi Nandakumar, Randall Bly, and Shyamnath
  Gollakota.
\newblock Detecting middle ear fluid using smartphones.
\newblock {\em Science Translational Medicine}, 11:eaav1102, 05 2019.

\bibitem{tam1}
Nam Bui, Nhat Pham, Jessica~Jacqueline Barnitz, Zhanan Zou, Phuc Nguyen, Hoang
  Truong, Taeho Kim, Nicholas Farrow, Anh Nguyen, Jianliang Xiao, Robin
  Deterding, Thang Dinh, and Tam Vu.
\newblock Ebp: An ear-worn device for frequent and comfortable blood pressure
  monitoring.
\newblock {\em Commun. ACM}, 64(8):118–125, jul 2021.

\bibitem{tymp}
Justin Chan, Ali Najafi, Mallory Baker, Julie Kinsman, Lisa Mancl, Susan
  Norton, Randall Bly, and Shyamnath Gollakota.
\newblock Performing tympanometry using smartphones.
\newblock {\em Communications Medicine}, 06 2022.

\bibitem{mobisys21}
Dong Ma, Andrea Ferlini, and Cecilia Mascolo.
\newblock Oesense: Employing occlusion effect for in-ear human sensing.
\newblock MobiSys '21, page 175–187, 2021.

\bibitem{eeg2}
Enea Ceolini, Jens Hjortkjær, Daniel Wong, James O'Sullivan, Vinay Raghavan,
  Jose Herrero, Ashesh Mehta, Shih-Chii Liu, and Nima Mesgarani.
\newblock Brain-informed speech separation (biss) for enhancement of target
  speaker in multitalker speech perception.
\newblock {\em NeuroImage}, 223:117282, 08 2020.

\bibitem{open-1}
Caslav Pavlovic, Volker Hohmann, Hendrik Kayser, Louis Wong, Tobias Herzke,
  S.~R. Prakash, zezhang Hou, and Paul Maanen.
\newblock Open portable platform for hearing aid research.
\newblock {\em The Journal of the Acoustical Society of America},
  143(3):1738--1738, 2018.

\bibitem{open-2}
T.~Herzke, H.~Kayser, F.~Loshaj, G.~Grimm, and V.~Hohmann.
\newblock Open signal processing software platform for hearing aid research (
  openmha ).
\newblock 2017.

\bibitem{paine2016fast}
Tom~Le Paine, Pooya Khorrami, Shiyu Chang, Yang Zhang, Prajit Ramachandran,
  Mark~A. Hasegawa-Johnson, and Thomas~S. Huang.
\newblock Fast wavenet generation algorithm.
\newblock 2016.

\bibitem{howard2017mobilenets}
Andrew~G. Howard, Menglong Zhu, Bo~Chen, Dmitry Kalenichenko, Weijun Wang,
  Tobias Weyand, Marco Andreetto, and Hartwig Adam.
\newblock Mobilenets: Efficient convolutional neural networks for mobile vision
  applications.
\newblock 2017.

\bibitem{ronneberger2015unet}
Olaf Ronneberger, Philipp Fischer, and Thomas Brox.
\newblock U-net: Convolutional networks for biomedical image segmentation.
\newblock 2015.

\bibitem{kingma2014adam}
Diederik~P Kingma and Jimmy Ba.
\newblock Adam: A method for stochastic optimization.
\newblock {\em arXiv preprint arXiv:1412.6980}, 2014.

\bibitem{voicecodecs}
Cox R. Neto S.F.de C.~Lamblin C. and Sherif M.H.
\newblock Itu-t coders for wideband, superwideband, and fullband speech
  communication.
\newblock IEEE, 2009.

\bibitem{timeslot}
Setting up the timeslot api.
  https://devzone.nordicsemi.com/nordic/short-range-guides/b/software-development-kit/posts/setting-up-the-timeslot-api,
  Jul 2015.

\bibitem{wireless-timesync}
Wireless timer synchronization among nrf5 devices.
  https://devzone.nordicsemi.com/nordic/short-range-guides/b/bluetooth-low-energy/posts/wireless-timer-synchronization-among-nrf5-devices,
  Jul 2016.

\bibitem{zhao2018sound}
Hang Zhao, Chuang Gan, Andrew Rouditchenko, Carl Vondrick, Josh McDermott, and
  Antonio Torralba.
\newblock The sound of pixels.
\newblock 2018.

\bibitem{luo2020endtoend}
Yi~Luo, Zhuo Chen, Nima Mesgarani, and Takuya Yoshioka.
\newblock End-to-end microphone permutation and number invariant multi-channel
  speech separation, 2020.

\bibitem{vctk}
Christophe Veaux, Junichi Yamagishi, Kirsten MacDonald, et~al.
\newblock Superseded-cstr vctk corpus: English multi-speaker corpus for cstr
  voice cloning toolkit.
\newblock 2016.

\bibitem{wham}
Gordon Wichern, Joe Antognini, Michael Flynn, Licheng~Richard Zhu, Emmett
  McQuinn, Dwight Crow, Ethan Manilow, and Jonathan~Le Roux.
\newblock Wham!: Extending speech separation to noisy environments.
\newblock {\em arXiv preprint arXiv:1907.01160}, 2019.

\bibitem{RISOUD2018259}
M.~Risoud, J.-N. Hanson, F.~Gauvrit, C.~Renard, P.-E. Lemesre, N.-X. Bonne, and
  C.~Vincent.
\newblock Sound source localization.
\newblock {\em European Annals of Otorhinolaryngology, Head and Neck Diseases},
  135(4):259--264, 2018.

\bibitem{allen1979image}
Jont~B Allen and David~A Berkley.
\newblock Image method for efficiently simulating small-room acoustics.
\newblock {\em The Journal of the Acoustical Society of America},
  65(4):943--950, 1979.

\bibitem{scheibler2018pyroomacoustics}
Robin Scheibler, Eric Bezzam, and Ivan Dokmani{\'c}.
\newblock Pyroomacoustics: A python package for audio room simulation and array
  processing algorithms.
\newblock In {\em 2018 ICASSP}, pages 351--355. IEEE.

\bibitem{sdr}
Jonathan~Le Roux, Scott Wisdom, Hakan Erdogan, and John~R. Hershey.
\newblock {SDR} - half-baked or well done?
\newblock {\em CoRR}, abs/1811.02508, 2018.

\bibitem{gutenberg}
Project gutenberg.
\newblock \url{https://www.gutenberg.org/}.
\newblock Accessed: 2021-12-20.

\bibitem{icaasp_dns}
Chandan K.~A. Reddy, Harishchandra Dubey, Vishak Gopal, Ross Cutler, Sebastian
  Braun, Hannes Gamper, Robert Aichner, and Sriram Srinivasan.
\newblock Icassp 2021 deep noise suppression challenge.
\newblock In {\em ICASSP 2021 - 2021 IEEE International Conference on
  Acoustics, Speech and Signal Processing (ICASSP)}, pages 6623--6627, 2021.

\bibitem{icaasp_dns_results}
https://www.microsoft.com/en-us/research/academic-program/deep-noise-suppression-challenge-interspeech-2021/.

\bibitem{DTLN}
Nils~L. Westhausen and Bernd~T. Meyer.
\newblock Dual-signal transformation lstm network for real-time noise
  suppression, arxiv, 2020.

\bibitem{sigsep}
Fabian-Robert Stöter, Antoine Liutkus, and Nobutaka Ito.
\newblock The 2018 signal separation evaluation campaign.
\newblock 2018.

\bibitem{wang2005ideal}
DeLiang Wang.
\newblock On ideal binary mask as the computational goal of auditory scene
  analysis.
\newblock In {\em Speech separation by humans and machines}, pages 181--197.
  Springer, 2005.

\bibitem{bt5}
{\em Bluetooth Core Specification v5.0}, 2016.

\bibitem{le-audio-faqs}
https://www.bluetooth.com/media/le-audio/le-audio-faqs.

\end{thebibliography}

\end{document}